\documentclass[11pt,letterpaper]{article}
\usepackage{fullpage}
\usepackage[T1]{fontenc}
\usepackage{lmodern,microtype,amsmath,amssymb,amsthm,mathtools}
\usepackage{booktabs,array,tabularx,enumitem,needspace,aliascnt}
\usepackage{xcolor}
\usepackage[sort]{natbib}
\usepackage[colorlinks=true,linkcolor=blue!45!black,citecolor=blue!45!black,urlcolor=blue!45!black]{hyperref}
\usepackage[capitalise,nameinlink]{cleveref}
\usepackage{makecell}

\newtheorem{theorem}{Theorem}[section]
\newaliascnt{lemma}{theorem}
\newtheorem{lemma}[lemma]{Lemma}
\aliascntresetthe{lemma}
\newaliascnt{corollary}{theorem}

\aliascntresetthe{corollary}
\theoremstyle{definition}
\newaliascnt{definition}{theorem}

\aliascntresetthe{definition}
\theoremstyle{remark}
\newaliascnt{remark}{theorem}

\aliascntresetthe{remark}
\crefname{lemma}{lemma}{lemmas}\Crefname{lemma}{Lemma}{Lemmas}
\crefname{corollary}{corollary}{corollaries}\Crefname{corollary}{Corollary}{Corollaries}
\crefname{definition}{definition}{definitions}\Crefname{definition}{Definition}{Definitions}
\crefname{remark}{remark}{remarks}\Crefname{remark}{Remark}{Remarks}

\newcommand{\cF}{\mathcal F}
\newcommand{\cS}{\mathcal S}
\newcommand{\bE}{\mathbb E}

\newcommand{\opt}{\operatorname{opt}}
\newcommand{\OPT}{\operatorname{OPT}}
\newcommand{\alg}{\operatorname{alg}}
\newcommand{\enc}{f}
\newcommand{\floorpowtwo}[1]{\lfloor#1\rfloor_{\mathrm{pow2}}}
\newcommand{\Daux}{D^{\mathrm{aux}}}
\newcommand{\excess}{E}

\newcommand{\ind}{\mathbf 1}

\setlist{itemsep=2pt,topsep=5pt}
\allowdisplaybreaks[1]
\title{Tight Bounds for Downward-Closed Prophet and Secretary Problems}
\author{
    Hanrui Zhang \\
    Chinese University of Hong Kong \\
    \texttt{hanrui@cse.cuhk.edu.hk}
}
\date{}

\begin{document}
\maketitle

\begin{abstract}
We study online selection problems subject to arbitrary downward-closed feasibility constraints, under different arrival orders, with different distributional assumptions.
We establish the following bounds: (1) $\Theta(\log^2 n / \log \log n)$ for the standard (adversarial-order) prophet problem, (2) $\Omega(\log n)$ for the (Bayesian, random-order) prophet secretary problem, and by extension, for the standard (prior-free, random-order) secretary problem, and (3) $O(\log n / \log \log n)$ for the free-order prophet problem.
Together with complementary bounds established in prior work, these results pin down the asymptotically tight competitive ratios for all these variants of prophet and secretary problems subject to downward-closed constraints.
Our results refine and reinforce the big-decisions-first principle by Rubinstein and Singla [2026], establishing a superconstant separation between the random-order and free-order models.
\end{abstract}

\tableofcontents
\clearpage

\section{Introduction}

{\em Prophet and secretary} problems are among the most natural and well studied online selection problems.
There are $n$ items arriving online, each with a value to be revealed upon arrival.
After observing the value, the algorithm must irrevocably make a decision to either accept or reject the item.
Accepted items are constrained: together, they must form a feasible set as specified by a given set system.
The reward collected by the algorithm is the total value of the accepted items, which is to be compared against the {\em prophet benchmark}, i.e., the maximum total value that an omniscient prophet can collect.
In particular, the typical goal is to optimize the {\em competitive ratio}, which is the ratio between the expectation of the prophet benchmark, and that of the algorithm's performance.
The set system that specifies feasibility deserves particular attention, since how rich it can be largely determines the difficulty of the problem.
In this paper, we consider {\em fully general} systems that are {\em downward-closed}.

Variants of the problem are induced by different arrival orders and distributional assumptions.
The following variants tend to receive the most attention:
\begin{itemize}
    \item The standard {\em prophet} problem: the items arrive in an adversarial order, and the algorithm knows the prior distributions of the items' (independent) values.
    \item The standard {\em secretary} problem: the items arrive in a uniformly random order, with no prior information about the values given to the algorithm.
    \item The {\em prophet secretary} problem: the items arrive in a uniformly random order, and the algorithm also knows the prior distributions of the values.
    \item The {\em free-order prophet} problem: the algorithm knows the prior distributions of the values, and even gets to choose an order of arrival.
\end{itemize}
The free-order prophet problem is no harder than the prophet secretary problem, which is no harder than both the standard prophet problem and the standard secretary problem.

With downward-closed feasibility systems, the asymptotically tight competitive ratios for all the above variants remained open, although the largest gap was already as small as $O((\log \log n)^2)$.
We close all these gaps.
In particular, we establish the following main results:
\begin{itemize}
    \item Lower bounds:
    \begin{itemize}
        \item $\Omega(\log^2 n / \log \log n)$ for the standard prophet problem.
        \item $\Omega(\log n)$ for the prophet secretary problem, which by extension also applies to the standard secretary problem.
    \end{itemize}
    \item Upper bounds:
    \begin{itemize}
        \item $O(\log n / \log \log n)$ for the standard prophet problem with binary values, which through the blackbox extension argument by \citet{Rub16} implies $O(\log^2 n / \log \log n)$ for the same problem with general values.
        \item $O(\log n / \log \log n)$ for the free-order prophet problem.
    \end{itemize}
\end{itemize}
For the standard prophet problem, our new bounds match.
For all other variants, our new bounds match the existing complementary bounds established in prior work.
The prior and new bounds are summarized in \Cref{tab:bounds}.

\begin{table}[t]
\centering\small
\setlength{\tabcolsep}{4pt}
\renewcommand{\arraystretch}{1.35}
\renewcommand{\tabularxcolumn}[1]{m{#1}}
\hypersetup{hidelinks}
\begin{tabularx}{\textwidth}{cccc}
\toprule
Model & Prior lower bound & Prior upper bound & Tight bound here\\
\midrule
Standard prophet
 & \makecell{$\displaystyle\Omega\!\left(\frac{\log^2 n}{(\log\log n)^2}\right)$ \\
   \citep{RS26}}
 & \makecell{$O(\log^2 n)$ \\ \citep{Rub16}}
 & \makecell{$\displaystyle\Theta\!\left(\frac{\log^2 n}{\log\log n}\right)$ \\
   Theorems~\ref{thm:adv-lower} and~\ref{thm:adv-upper}}\\
\addlinespace
Prophet secretary
 & \makecell{$\displaystyle\Omega\!\left(\frac{\log n}{\log\log n}\right)$ \\ 
   \citep{BIK07}}
 & \makecell{$O(\log n)$ \\ \citep{RS26}}
 & \makecell{$\Theta(\log n)$ \\ Theorem~\ref{thm:ps-lower}}\\
\addlinespace
Standard secretary
 & \makecell{$\displaystyle\Omega\!\left(\frac{\log n}{\log\log n}\right)$ \\ 
   \citep{BIK07}}
 & \makecell{$O(\log n)$ \\ \citep{RS26}}
 & \makecell{$\Theta(\log n)$ \\ Theorem~\ref{thm:ps-lower}}\\
\addlinespace
Free-order prophet
 & \makecell{$\displaystyle\Omega\!\left(\frac{\log n}{\log\log n}\right)$ \\ 
   \citep{BIK07}}
 & \makecell{$O(\log n)$ \\ \citep{RS26}}
 & \makecell{$\displaystyle\Theta\!\left(\frac{\log n}{\log\log n}\right)$ \\ 
   Theorem~\ref{thm:free-upper}}\\
\bottomrule
\end{tabularx}
\caption{Competitive ratios for arbitrary downward-closed feasibility systems.}
\label{tab:bounds}
\end{table}

Conceptually, these tight bounds refine and reinforce the {\em big-decisions-first} principle of \citet{RS26}: the free-order problem admits the smallest tight ratio, followed by the random-order problems (where distributional information does not even help asymptotically), followed by the adversarial-order problem.
This corresponds to increasingly little freedom for the algorithm to process ``bigger'' items first.\footnote{
    This is best illustrated in the lower bound constructions, where all items are weighted Bernoulli, and the weight measures how ``big'' an item is.
}
\citet{RS26} already establish the separation between adversarial-order and random-order models, which is further tightened in this paper.
In addition, we establish a smaller but superconstant separation between random-order and free-order models too.

\subsection{Technical Overview}

\subsubsection{Lower Bounds}

We first discuss the lower bound constructions, starting with a common structure that unifies both our new lower bounds and the previous almost tight lower bound by \citet{RS26}.

\paragraph{A shared structure for lower bounds: layered self-encoding.}
Consider the following abstract construction.
The entire instance is divided into $T$ {\em layers} (where typically $T = \Theta(\log n)$, generally indexed by $t$.
We generally construct the prophet benchmark by visiting the layers in the order of $1$ to $T$.
In the adversarial-order model, we (as well as \citet{RS26}) force the algorithm to process the layers in the reverse order of $T$ to $1$.
By the big-decisions-first principle, one would naturally expect items in earlier layers to be bigger.
In our constructions, all items in a single layer are independently activated with the same probability, and each layer contributes roughly the same amount in expectation to the prophet benchmark.
Accordingly, later layers host more items (which are smaller) than earlier layers.

Each layer $t$ is further divided into $d_t$ {\em coordinates}.
These coordinates contain disjoint items, and all items within a single coordinate are mutually exclusive, i.e., a feasible selection can contain at most one item within a single coordinate.
Items in each coordinate are indexed by two ``symbols'': one from a ``message'' alphabet $\Sigma_M$, and one from a ``codeword'' alphabet $\Sigma_{C, t}$ which can vary across layers (we will explain what messages and codewords are momentarily).
So within a coordinate in layer $t$, there are $|\Sigma_M| \cdot |\Sigma_{C, t}|$ mutually exclusive items.
And layer $t$ contains $d_t \cdot |\Sigma_M| \cdot |\Sigma_{C, t}|$ items in total.
We will index items using the tuple $(t, i, (m_t)_i, (c_t)_i)$, i.e., the item in coordinate $i$ of layer $t$, whose corresponding message symbol and codeword symbol are $(m_t)_i$ and $(c_t)_i$.
One caveat is that items here might be conceptual: in the adversarial-order construction, we actually replace each conceptual item with a group of identical physical items that are jointly feasible, but each incompatible with any other groups within the same coordinate (so the groups inherit the feasibility of the original conceptual items).
Physical items within a group need not be explicitly indexed (at least not in this overview): the groups simply force the same effect in the mutually exclusive groups construction by \citet{BIK07} that forces the algorithm to commit to a group as soon as it takes one physical item in that group.
This contributes a factor of $\Omega(\log n / \log \log n)$ in the adversarial-order lower bound.
From now on, we will use the term ``items'' by default, and say ``groups'' in the adversarial-order construction, where we need to distinguish between groups and physical items.

To describe the feasibility system, we will adopt a self-coding interpretation.
Each layer $t$ carries a message $m_t$, whose length is $d_t$ (so each coordinate of the layer is also a coordinate of the message), i.e., $m_t \in \Sigma_M^{d_t}$.
Each layer also ``inherits'' a codeword $c_t$ whose length is also $d_t$, which is conceptually obtained by encoding messages in previous layers $m_1, \dots, m_{t - 1}$ using a public encoder $\enc_t$, i.e., $c_t = \enc_t(m_1, \dots, m_{t - 1}) \in \Sigma_{C, t}^{d_t}$.
The encoder is expected to have some kind of error correction property: on two different input sequences of messages, the output codes can only intersect on few coordinates.\footnote{
    Parts of the input messages may be dropped before encoding, in which case differences in the dropped part do not force small intersections.
    This happens in our prophet secretary construction.
}
This crucial property forces the online algorithm to make commitments that greatly limits its subsequent options --- we will get back to this point soon.

Each maximal feasible set is specified by a sequence of messages $(m_1, \dots, m_T)$, one for each layer.
Let $c_t = \enc_t(m_1, \dots, m_{t - 1})$ for each $t$.
The corresponding set of items is
\[
    \{(t, i, (m_t)_i, (c_t)_i) \mid t \in [T], i \in [d_t]\}.
\]
That is, in every layer, we collect the item in each coordinate whose two symbols match the corresponding coordinates of $m_t$ and $c_t$.
Each layer $t$ contributes precisely $d_t$ conceptual items.

Intuitively, the layered self-encoding structure gives the prophet greater freedom, because in the order of $1$ to $T$, upon processing each fresh layer, the prophet gets to choose the entire message $m_t$.
Our constructions make it so that the prophet can extract a constant fraction of the maximum value possible in this layer.
The algorithm, on the other hand, is forced to either process items from different layers in a uniformly random order, or from layer $T$ to layer $1$.
So the algorithm inevitably faces a dilemma:
\begin{itemize}
    \item If the algorithm takes too many items from later (according to the prophet order) layers, it must have already committed to a codeword in some layer $t$, because error correction ensures that two different codewords cannot contain too many items in common in one layer.
    This pins down (a large part of) the messages in previous layers $m_1, \dots, m_{t - 1}$, and by implication also the codewords in all previous layers $c_1, \dots, c_{t - 1}$.
    As a result, most (if not all) choices in previous layers are fixed, which forces a loss compared to the flexible prophet benchmark.
    \item If the algorithm waits for items in earlier layers and takes only a few items in later layers in the meantime, then it is already missing a lot of value in the process of waiting, which is also a loss compared to the prophet benchmark.
\end{itemize}

Our new constructions, as well as the earlier construction by \citet{RS26}, are based on this intuitive dilemma.
They apply different concrete constructions (especially in the encoder $\enc_t$) in response to different arrival orders, and use different accounting arguments to bound the inevitable losses.
We discuss them next.

\paragraph{The standard prophet lower bound: a code sensitive to first divergence.}
Under the layered self-encoding interpretation, our lower bound construction for the adversarial-order model can be directly compared with that by \citet{RS26}.
The main differences are twofold: the code being used, and how same-layer hardness is exploited.
We first review the construction by \citet{RS26}.
Roughly speaking, the code used by \citet{RS26} has the following guarantee: for each layer $t$, sequences of messages $(m_1, \dots, m_t)$ and $(m_1', \dots, m_t')$, and the corresponding codewords $c_t = \enc_t(m_1, \dots, m_t)$ and $c_t' = \enc_t(m_1', \dots, m_t')$,\footnote{
    \citet{RS26} allow the codeword $c_t$ to depend on $m_t$ in addition to $m_1, \dots, m_{t - 1}$.
    Superficially, this goes only slightly beyond the layered self-encoding framework.
    But it allows their code to enforce same-layer hardness, which is handled by replacing conceptual items with groups of physical items in our construction.
    This is one of the two key differences that is less relevant to our construction.
}
\begin{itemize}
    \item If $m_t \ne m_t'$, then $|c_t \cap c_t'| / d_t \le 1 / \Theta(T)$.
    \item If $m_\tau \ne m_\tau'$ for some $\tau < t$, then $|c_t \cap c_t'| / d_t \le 1 / \Theta(T^2)$.
\end{itemize}
Here, $d_t$ is the number of coordinates in layer $t$, $T = \Theta(\log n / \log \log n)$ is the number of layers in their construction, and $|c_t \cap c_t'| = |\{i \mid (c_t)_i = (c_t')_i\}|$ is a (notation-abusing) shorthand for the number of coordinates in which $c_t$ and $c_t'$ agree.

To see why this forces a large loss by the online algorithm, imagine its execution from layer $T$ to layer $1$.
As soon as the algorithm takes more than $\Theta(d_t / T^2)$ items in a layer $t$, it has effectively committed to a sequence of message $m_1, \dots, m_{t - 1}$ up to layer $t - 1$, because two different sequences cannot both contain that many items in layer $t$.
We call this the phase transition.
Before the phase transition, the algorithm can only take an $O(1 / T^2)$ fraction of items per layer, losing the same factor against the prophet benchmark.
In the layer where the transition happens, the algorithm can freely choose up to an $O(1 / T)$ fraction of the items taken by the prophet benchmark.\footnote{
    Taking more would commit to the message in the current layer too, which means all future items in the layer are fixed and subject to discounting because of rare activations.
    These items therefore contribute little value.
}
This is a smaller loss, but it only happens in one layer, so it makes the algorithm better off only by a constant factor.
Finally, after the tranisition, the algorithm has committed to a particular sequence of messages, which means the best it can do is to take all items corresponding to that sequence in all remaining layers.
The activation probability of each item is set to $\Theta(1 / T^2)$, which again forces a loss of the same factor.\footnote{
    The construction of \citet{RS26} does not replace conceptual items by groups of physical items.
    Such replacements would not immediately strengthen their construction since the code in their construction already incorporates same-layer hardness that the replacements are supposed to capture.
}
Tuning the parameters gives $T = \Theta(\log n / \log \log n)$, and a lower bound of $\Omega((\log n / \log \log n)^2)$.

We use a different code that is sensitive to the ``first divergence'' in a finer way.
For each layer $t$, different sequences of messages $(m_1, \dots, m_{t - 1})$ and $(m_1', \dots, m_{t - 1}')$, and the corresponding codewords $c_t = \enc_t(m_1, \dots, m_{t - 1})$ and $c_t' = \enc_t(m_1', \dots, m_{t - 1}')$, our code guarantees that $|c_t \cap c_t'| = O(d_t / \floorpowtwo{t - \tau}^2)$, where $\tau < t$ is the smallest index where $m_\tau \ne m_\tau'$, and $\floorpowtwo{x}$ is the largest power of $2$ that is no larger than $x$.
Note that this is not a ``stronger'' code than that of \citet{RS26}: in fact, when $t - \tau$ is small, their code guarantees a much smaller overlap between $c_t$ and $c_t'$, which generally helps the lower bound construction.
Nonetheless, taking a more global perspective, what really matters is the {\em sum} of these overlaps in all layers since the first divergence between two sequences of messages.
Assuming the prophet gains $1$ per layer, the algorithm's performance after the first divergence is bounded by (a constant multiple of)
\[
    1 + 2 \times 1 / 2^2 + 4 \times 1 / 4^2 + 8 \times 1 / 8^2 \dots = \sum_{j \ge 0} 2^{-j} = 2.
\]
In other words, the total gain of the algorithm after the first divergence is only about the prophet's gain in one layer, which is a loss of a factor $\Theta(T)$, where $T = \Theta(\log n)$ is the number of layers in our construction.
In layers smaller than the first divergence,\footnote{
    The first divergence here is retrospective and informal.
    The actual execution of the algorithm instead progressively limits possible messages, until the sequence of messages up to the layer being processed is entirely fixed.
    This is what we call the first divergence.
}
the algorithm cannot choose the messages any more, so it suffers the inevitable loss of order $\Theta(T)$.

The above argument only contributes one of two factors in our lower bound.
The other factor comes from replacing conceptual items with groups of physical items, each of size $\Theta(\log n / \log \log n)$.
This forces a loss of the same factor on top of the $\Theta(T)$ factor given by the first divergence argument, because the prophet chooses among many independent groups at each coordinate and finds an all-active group with constant probability.\footnote{
    Importantly, this is possible because upon reaching each layer $t$, the prophet is free to choose the message $m_t$ without any constraints.
    In the construction by \citet{RS26}, this choice is coupled with the codeword in the same layer.
}
The algorithm, on the other hand, must commit to a group as soon as it chooses the first physical item in it.
Tuning the parameters gives an overall bound of $\Theta(\log^2 n / \log \log n)$.

\paragraph{The prophet secretary lower bound: nested prefix encoding.}
Our construction for the random-order model uses a very different code tailored to the specific arrival order.
The code itself is actually simpler and quite standard, except for the part where it neglects parts of the input sequence of messages $(m_1, \dots, m_{t - 1})$, which we explain first.
Suppose the size of the alphabet for messages $|\Sigma_M| = 2^T$, and every coordinate $(m_\tau)_i$ of a message is represented as a binary string of length $T$.
For every $\tau < t$, the encoder $\enc_t$ in layer $t$ depends only on the first $t - \tau$ bits of each $(m_\tau)_i$.
Taking another perspective, fixing $\tau$ and letting $t$ grow, $\enc_t$ encodes a nested sequence of prefixes of each $(m_\tau)_i$ that also grows.
Importantly, for two layers $t < t'$, $\enc_t$ encodes a strict subset of the information encoded by $\enc_{t'}$, so knowing $c_{t'} = \enc_{t'}(m_1, \dots, m_{t' - 1})$ also uniquely determines $c_t = \enc_t(m_1, \dots, m_{t - 1})$.

Restricted to the part of the messages that is actually encoded, the code gives a fairly standard guarantee (provided by, e.g., the Reed-Solomon code): for any layer $t$ and sequences of messages $(m_1, \dots, m_{t - 1})$ and $(m_1', \dots, m_{t - 1}')$ that differ on the parts actually encoded by $\enc_t$, $|c_t \cap c_t'| = O(d_t / T)$, where $c_t = \enc_t(m_1, \dots, m_{t - 1})$ and $c_t' = \enc_t(m_1', \dots, m_{t - 1}')$.
In other words, as soon as the algorithm accepts more than $\Theta(d_t / T)$ items in a layer $t$, the codeword $c_t$ is uniquely determined, which means:
\begin{itemize}
    \item The codeword $c_\tau$ for any $\tau < t$ is also uniquely determined.
    \item The encoded prefixes of $m_1, \dots, m_{t - 1}$ are also determined.
\end{itemize}
The second point in particular allows decreasing freedom for smaller layers $\tau$ as $t - \tau$ becomes larger.
Importantly, we have to achieve the guarantee using an alphabet for the codeword $\Sigma_C$ (we omit $t$ in the subscript because this alphabet does not depend on $t$ in this construction) that is as small as possible.
This is optimized in the choice of parameters.

Now we discuss the actual lower bound argument.
Throughout the execution of the algorithm, we track an index $u$, which is the largest layer in which the algorithm has accepted more than $\Theta(d_u / T)$ items to force a unique codeword in layer $u$.
Call this the cutoff layer.
We divide items accepted by the algorithm into two types:
\begin{itemize}
    \item Unconstrained items: the item lies in layer $t$, and immediately before the item's value is revealed, the cutoff layer $u < t$.
    In other words, the determined codeword $c_u$ itself does not constrain the acceptance decision.
    \item Constrained ones: the item lies in layer $t$, and immediately before the item's value is revealed, the cutoff layer $u \ge t$.
    In other words, $c_u$ partially constrains the acceptance decision, because (1) the codeword $c_t$ in layer $t$ is already fixed given $c_u$, and (2) we have to fix some coordinate of $m_t$ (a binary string of length $T$) to accept the item, which must be consistent with the prefix dictated by $c_u$.
\end{itemize}
Observe that unconstrained items cannot contribute much anyway: the number of such items in layer $t$ is at most $O(d_t / T)$, whereas the prophet accepts $\Theta(d_t)$ activated items in the same layer in expectation, leaving a gap of $\Omega(T)$.

The constrained ones require more attention.
Intuitively, the constrained ones cannot contribute much too: suppose, for simplicity, that the algorithm first accepts only unconstrained items, and after that it transitions into only accepting constrained items.
Then, the total value contributed by constrained items cannot be too large.
This is because the construction sets the activation probability of each item to $1 / 2^T$, so once the codeword $c_t$ in a layer $t$ is fixed, there is, in expectation, precisely $1$ item in each coordinate $i$ that is activated (the prophet of course collects this item whenever it exists).
On the other hand, from the algorithm's perspective, a prefix of length $u - t$ is already fixed, so the algorithm only gets to choose from $2^{T - u + t}$ items in each coordinate, with an expected gain of at most $2^{t - u}$ in the entire layer ($2^{t - u} / d_t$ in expectation in each coordinate, summed over $d_t$ coordinates).
This is true even assuming the algorithm can take all items consistent with the fixed prefix, ignoring other feasibility constraints.
Further summing over $t \le u$, this gives us $\sum_{t \le u} 2^{t - u} = O(1)$, establishing a gap of $\Omega(T)$.
The nested prefix encoding is just strong enough to enable this very sum.

In general, the algorithm of course is allowed to interleave the two types of acceptances.
This creates additional technical difficulties.
To handle them, the key observation is that the cutoff layer $u$ increases monotonically throughout the execution of the algorithm, and it can strictly increase only $T$ times.
This gives us an easy way to essentially union bound over the possible behavior of the algorithm and preserve the above intuition.
We interpret the uniformly random arrival order in the following way: there are $T$ bins $B_1, \dots, B_T$.
Each item is first assigned independently to a uniformly random bin, and then each bin is internally permuted uniformly at random.
Items in a single bin arrive consecutively, and bins are ordered from $1$ to $T$.

We relax the algorithm with the following additional power: it observes the entire order of items, including the bin assignments, before items arrive.
In addition, we focus on constrained acceptances and ignore unconstrained ones (which are handled separately above), and we make the cutoff layer $u$, together with the determined codeword $c_u$, something the algorithm can choose freely subject to monotonicity.
Initially, $u = 0$.
Upon entering a new bin, the algorithm can choose to strictly increase $u$ any number of times; at each increase, it can even choose the determined codeword $c_u$ arbitrarily (i.e., it need not be consistent with the algorithm's previous choices for smaller $u$).
As soon as the algorithm enters a bin, as well as when an increase of $u$ happens, the algorithm immediately harvests all items in the current bin in layers $t \le u$ that are consistent with $c_u$, even if these items are mutually incompatible.
The actual value of an item can only be harvested once, even if it appears in two harvests.
We call this the harvesting game.
Below we argue why it still gives a lower bound of $\Omega(T)$ even with such additional freedom.

We first bound the intersection of the items compatible with a codeword $c_u$ (call it the {\em menu} $M(c_u)$) and a particular bin $B_j$.
We do this by examining smaller components of menus, which we call {\em pieces}.
A piece $P_{t, i, s, p}$ is defined by a layer $t$, a coordinate $i$, a codeword symbol $s \in \Sigma_C$, and a prefix $p \in \{0, 1\}^\ell$ of a message symbol of length $\ell \le T$ (recall that each message symbol is a binary string of length $T$).
It contains all items in coordinate $i$ of layer $t$, whose codeword symbol is $s$, and whose message symbol has the prefix $p$.
Every menu is the union of disjoint pieces, and one can show there are $O(n)$ possible pieces in total.
We apply a concentration bound on the size of the intersection between every piece and every bin, followed by a union bound: with high probability, every piece $P_{t, i, s, p}$ intersects every bin $B_j$ on $O(|P_{t, i, s, p}| / T + T)$ items.
The additive $O(T)$ contributes very little to a menu even aggregated over all pieces.
This means essentially the same concentration happens for all menus simultaneously with high probability.
We condition on this event (the complementary case happens too rarely and cannot contribute much).
The rest of the argument is simple: the algorithm performs $O(T)$ harvests in total (one entering each bin, and one for each increase of $u$), and in each harvest, it collects $O(1 / T)$ of the value available in $M(c_u)$ in expectation, which is $O(1 / T)$. The total value harvested is therefore $O(1)$, creating a gap of $\Omega(T)$.

\subsubsection{Upper Bounds}

Now we discuss the upper bounds, both of which involve an induction on the moment generating function of the prophet benchmark that connects it to the optimal online algorithm.
We first discuss this induction instantiated for binary values, and how it leads to a bound of $O(\log n / \log \log n)$ irrespective of the arrival order.
The proof compares the recurrence satisfied by the optimal online algorithm with a similar recurrence involving the moment generating function of the prophet benchmark.
The argument crucially depends on observing (or really, guessing) the right scale of comparison.

\paragraph{The binary values upper bound: a comparison between two recurrences.}
Suppose $\cF$ is the feasibility system, and $X_i$ is the Bernoulli random value of each item $i$.
For any (residual) set system $\cS$, let $\OPT(\cS) = \max_{S \in \cS} \sum_{i \in S} X_i$ denote the (random) prophet benchmark on $\cS$.
$\OPT = \OPT(\cF)$ is the standard prophet benchmark.
We define inclusions and exclusions of an item $i$ in a system $\cS$:
\begin{itemize}
    \item $\cS$ including $i$ $\cS / i = \{S \setminus \{i\} \mid S \in \cS, i \in S\}$.
    \item $\cS$ excluding $i$ $\cS \setminus i = \{S \setminus \{i\} \mid S \in \cS\}$.
    This is essentially $\cS$ restricted to all items other than $i$.
\end{itemize}
For a feasible item $i$, $\OPT(\cS)$ satisfies the following recurrence:
\[
    \OPT(\cS) = \max\{\OPT(\cS \setminus i), X_i + \OPT(\cS / i)\}.
\]
Based on this recurrence, restricted to binary values, one can show that for any $a \ge 2$,
\[
    \bE[a^{\OPT(\cS)}] \le \bE[a^{\OPT(\cS \setminus i)}] + p_i (a - 1) \bE[a^{\OPT(\cS / i)}],
\]
where $p_i$ is the probability that $X_i = 1$.
We will compare this with a recurrence for the optimal algorithm that has a similar form.

Consider a fixed adversarial order (the proof in fact works for adaptive adversarial orders that depend only on observed values by the algorithm).
Let $\alg(\cS)$ be the expected (onward) value of the optimal online algorithm on $\cS$. Suppose item $i$ arrives next.
Observe that $\alg(\cS \setminus i) \ge \alg(\cS / i)$.
So we have the following simple recurrence:
\begin{equation}
    \alg(\cS) = \alg(\cS \setminus i) + p_i \cdot \max\{0, 1 + \alg(\cS / i) - \alg(\cS \setminus i)\}. \label{eq:alg_comparison}
\end{equation}
We will inductively argue that
\[
    \bE[a^{\OPT(\cS)}] \le \exp(O(a + \log m) \cdot \alg(\cS)),
\]
where $m$ is the number of items involved in $\cS$, for any $a \ge 2$ by comparing the two recurrences.
The essence of the entire argument lies in {\em identifying} the above bound, which is essentially done by reverse engineering from the conclusion that we need it to lead to: by choosing $a = \Theta(\log n)$ and $\cS = \cF$, this (together with convexity) immediately implies
\[
    \Theta(\log \log n) \cdot \bE[\OPT(\cF)] = \log \left(a^{\bE[\OPT(\cF)]}\right) \le \log \bE[a^{\OPT(\cF)}] \le \Theta(\log n) \cdot \alg(\cF),
\]
which means
\[
    \alg(\cF) \ge \Theta(\log \log n / \log n) \cdot \bE[\OPT(\cF)].
\]

Coming back to the inductive proof of the bound on the moment generating function, the induction hypothesis gives
\begin{align*}
    \bE[a^{\OPT(\cS)}] & \le \bE[a^{\OPT(\cS \setminus i)}] + p_i (a - 1) \bE[a^{\OPT(\cS / i)}] \\
    & \le \exp(O(a + \log (m - 1)) \cdot \alg(\cS \setminus i)) + p_i (a - 1) \cdot \exp(O(a + \log (m - 1)) \cdot \alg(\cS / i)).
\end{align*}
We only need to bound the latter by $\exp(O(a + \log m) \cdot \alg(\cS))$, which can be expanded by \Cref{eq:alg_comparison} to provide a clearer comparison.
We identify two complementary cases: when $1 + \alg(\cS / i) - \alg(\cS \setminus i) > 1/2$, i.e., when an activated item brings relatively large gain to the algorithm, and when the same quantity is at most $1/2$.
In the two cases, different quantities dominate, which give the desired inequality in both cases.
Technical details are omitted.
We remark that the proof relies crucially on binary values in several places, and does not appear easily generalizable to unrestricted values.
Our upper bound for the free-order model takes a different approach.

\paragraph{The free-order prophet upper bound: approximation by activated exponentials.}
The free-order bound has a clean high-level structure:
We first identify a special class of variables, namely {\em activated exponentials}.
An activated exponential variable of scale $s$ is the product of a Bernoulli variable (the activation), and an exponential variable of mean $s$, independent of each other.
A similar inductive argument to the one above shows that, when the values of items are activated exponentials, if we visit the items in decreasing order of the scale, then the optimal online algorithm loses only a factor of $2$ compared to the prophet benchmark.
We will omit a detailed discussion of this part, since it is structurally very similar to the argument demonstrated above.

Now for general variables, we present a reduction to activated exponential variables, which loses a factor of $\Theta(\log n / \log \log n)$.
The argument handles each item $i$ separately.
It constructs $\Theta(\log n / \log \log n)$ auxiliary random variables $Y_{i, j}$, all being activated exponentials.
Moreover, a distribution $\Daux_i$ is specified over these auxiliary random variables.
The desired guarantee is item-wise mutual domination within a margin of $\Theta(\log n / \log \log n)$: each $Y_{i, j}$ is dominated by $X_i$ in terms of the reward it generates (more detail to follow), for both the prophet and the online algorithm; all auxiliary variables $Y_{i, j}$ on average, amplified by a factor of $\Theta(\log n / \log \log n)$, dominate the original variable $X_i$.
Additional reductions are omitted here, which cost $O(k)$ overall.
We next discuss how this (combined with the factor-$2$ argument for activated exponentials) directly leads to the desired bound.

The actual algorithm is the following: for each $X_i$, we sample some $j_i$ from $\Daux_i$, and let $s_i$ be the scale of $Y_{i, j_i}$.\footnote{
    The actual construction first normalizes each $X_i$ such that its mean becomes $1$.
    It then runs this construction, and obtain the true scale by multiplying $s_i$ with the original mean of $X_i$.
    We ignore this for simplicity here.
}
We visit items in decreasing order of $s_i$, and run the actual optimal online algorithm for $X_i$ in that order.
Abusing notation, let $\OPT(\{X_i\})$ and $\OPT(\{Y_{i, j_i}\})$ denote the prophet benchmark on the respective variables, and $\alg_\sigma(\{X_i\})$ and $\alg_\sigma(\{Y_{i, j_i}\})$ be the optimal algorithm's (expected) performance fixing a particular order $\sigma$.
The above algorithm works because fixing $\{j_i\}$ (and therefore $\sigma$),
\[
    \alg_\sigma(\{X_i\}) \ge \alg_\sigma(\{Y_{i, j_i}\}),
\]
because of domination of every $Y_{i, j_i}$ by $X_i$.
On the other hand, averaging over the choices of $\{j_i\}$ (and therefore $\sigma$), we have
\[
    \bE[\alg_\sigma(\{Y_{i, j_i}\})] \ge \bE[\OPT(\{Y_{i, j_i}\}) / 2] \ge \Omega(\log \log n / \log n) \cdot \bE[\OPT(\{X_i\})],
\]
because of domination on average of $\{X_i\}$ by $\{Y_{i, j_i}\}$ up to a factor of $O(\log n / \log \log n)$.
Putting the two inequalities together, we get
\[
    \bE[\alg_\sigma(\{X_i\})] \ge \Omega(\log \log n / \log n) \cdot \bE[\OPT(\{X_i\})],
\]
which is the desired bound.

Now we zoom into a single item $i$, and consider the construction of $Y_{i, j}$.
Without loss of generality, suppose $\bE[X_i] = 1$.
Then, by standard reductions, the range that matters in a nontrivial way is $[1, 2n]$.
We divide this range into $k = \Theta(\log n / \log \log n)$ geometric intervals, whose endpoints are $1 = x_1 < x_2 < \dots < x_k$.
In particular, $x_{j + 1} = k x_j$, and $x_k \ge 2n$.
We will associate $Y_{i, j}$ with the interval $[x_j, x_{j + 1}]$.
In particular, the contribution of $Y_{i, j}$ alone (properly weighted and scaled) already dominates $X$ up to a constant factor on this interval.
So, each interval is covered by at least one auxiliary variable.

Our notion of domination is based on the excess function $\excess_X(t) = \bE[\max\{X - t, 0\}]$ for a random variable $X$ and a given threshold $t$.
In particular, we say $X$ dominates $X'$ iff $\excess_X(t) \ge \excess_{X'}(t)$ for all $t \ge 0$.
Both the performance of the optimal online algorithm and the prophet benchmark are monotone under this notion of domination, which enables the reduction to activated exponentials discussed above.
Suppose the weight we give $Y_{i, j}$ is $w_{i, j}$.
The probability that we sample $Y_{i, j}$ is $w_{i, j} / W_i$, where $W_i = \sum_j w_{i, j}$.
The desired dominations are:
\begin{itemize}
    \item For each $j$ and $t \ge 0$: $\excess_{Y_{i, j}}(t) \le \excess_{X_i}(t)$, plus some small shift.
    \item For each $j$ and $t \in [x_j, x_{j + 1})$: $(w_{i, j} / W) \cdot \excess_{O(k) \cdot Y_{i, j}}(t) \ge \excess_{X_i - O(k)}(t)$.
\end{itemize}
These guarantees are achieved using a delicate but well motivated construction of both $\{Y_{i, j}\}$ and $\{w_{i, j}\}$.
Details are omitted here.

\subsection{Further Related Work}

The classical secretary problem and prophet inequality originate in optimal stopping: a single irrevocable choice is compared with the best choice in hindsight \citep{Dyn63,KS77,KS78}. \citet{HKS07} observe their connections to auctions and pricing.
See \citep{Luc17} for the economic perspective, \citep{CFH18} for a more focused overview on single-item variants, and \citep{GS21} for more on random-order models.
We discuss several particularly relevant lines of results below.

\paragraph{General downward-closed systems.}
For general downward-closed systems, the disjoint-groups example by \citet{BIK07} yields an $\Omega(\log n/\log\log n)$ obstruction already for binary values.
In a major breakthrough, \citet{Rub16} gives $O(\log n)$ guarantees for binary prophet and secretary problems, and a blackbox reduction giving $O(\log n \log r) = O(\log^2 n)$ for general values, where $r$ is the rank of the feasibility system.
\citet{RS26} establish an $\Omega((\log n/\log\log n)^2)$ prophet lower bound and an $O(\log n)$ secretary upper bound, closing the gap up to $\mathrm{poly} \log \log$ factors. Our adversarial-order result improves both sides of the former gap, while our prophet-secretary lower bound matches the latter upper bound. In particular, our binary-value upper bound removes an $O(\log \log n)$ factor from the $O(\log n)$ binary prophet guarantee of \citet{Rub16}.

\paragraph{Matroid prophet and secretary problems.}
Matroids are a natural and well-studied class of feasibility systems.
\citet{BIK07} introduce the matroid secretary problem, and also discuss the general downward-closed variant. For general matroids, the sequence of work from \citep{CL12} to \citep{Lac14} and \citep{FSZ15} improves the secretary guarantee to $O(\log\log r)$, where $r$ is the rank.
The very recent results of \citep{Singla26,Huang26,Abdi26} subsequently push the ratio to a constant, and eventually to the optimal $e$. On the Bayesian side, \citet{KW12} establish the tight factor-two matroid prophet inequality. Online contention resolution provides a related framework for converting fractional solutions into feasible online selections \citep{FSZ16}.
\citet{DFKL20} give prophet inequalities for a broad range of combinatorial constraints, including matroids, through posted prices.
These results exploit structural properties, such as exchange, that arbitrary downward-closed families need not satisfy.

\paragraph{Prophet secretary and related problems.}
\citet{Esfandiari15} introduce the prophet secretary problem, which assumes both uniformly random arrival orders and known independent prior distributions.
This is weakly easier than both prophet and secretary.
\citet{azar2017prophetsecretarysurpassing11e} show that the optimal ratio for this problem is strictly better than $1 - 1/e$.
\citet{CSZ19} study blind strategies for the problem.
For richer (but structured) feasibility, \citet{EHKS24} study prophet secretary for matroids and combinatorial auctions. \citet{JSZ13} also study a free-order secretary model subject to matroid feasibility, which changes the order model without supplying known independent value distributions.

\paragraph{Combinatorial objectives and stochastic probing.}
\citet{RS17} study prophet inequalities for combinatorial objectives, and the secretary result of \citet{RS26} extends beyond additive objectives to XOS valuations.
Combinatorial auctions are another important model for prophet inequalities: \citet{FGL15} study posted-price mechanisms for combinatorial auctions; \citet{DFKL20} present a unified framework that covers combinatorial auctions; \citet{CC23} present constant-factor prophet inequalities for subadditive valuations in particular.
Stochastic probing separates inspection feasibility from final selection feasibility: \citet{GN13} formulate this model, \citet{GNS17} study submodular and XOS adaptivity gaps, and \citet{LLZ25} give bounds for subadditive functions.

\paragraph{Sample-based prophet inequalities.}
\citet{AKW14} systematically study prophet inequalities with limited sample access.
\citet{RWW20} present optimal single-choice prophet inequalities from samples.
\citet{CDF22} study greedy-ordered single-sample approaches.
\citet{Fu24} and \citet{FSZ26} study sample-based prophet inequalities and online contention resolution for matroids.

\section{Preliminaries}

There are $n$ items and a known nonempty downward-closed feasibility system
$\cF\subseteq2^{[n]}$: every subset of a feasible set is feasible. On observing an
item's nonnegative value, the algorithm must irrevocably accept or reject it,
and its accepted items must remain in $\cF$. For a value vector $x$, write
\[
 \OPT(\cF;x)=\max_{S\in\cF}\sum_{i\in S}x_i.
\]
This is the (realized) prophet benchmark under $x$.
We suppress $\cF$ when it is fixed. We allow unrestricted computation.

\paragraph{Standard prophet and free-order prophet.}
The values $X_1,\ldots,X_n$ of the $n$ items are independent random variables with
known distributions. The expected prophet benchmark is $\opt=\bE[\OPT(\cF;X)]$. In the
standard prophet problem the arrival order is fixed adversarially; in the
free-order prophet problem the algorithm chooses the order from the priors,
before observing the values (our upper bound uses non-adaptive orders). For a fixed order $\sigma$, let
$\alg_\sigma$ denote the optimal expected {\em online} reward. We write $\alg$ when
the order is clear. A competitive ratio of $R$ means $\alg\ge\opt/R$ on every
instance.

\paragraph{Standard secretary and prophet secretary.}
In both models the arrival order is a uniformly random permutation. In the
standard secretary problem the values $x_i$ are fixed adversarially and not
known in advance; here $\opt=\OPT(\cF;x)$ and the guarantee
$\alg\ge\opt/R$ must hold for each value vector, with expectation over the
order and the algorithm's randomness. In prophet secretary the values have
known independent distributions, the permutation is independent of them, and
$\opt=\bE[\OPT(\cF;X)]$; the algorithm's expected reward also averages over the
values. Thus $\opt$ and $\alg$ always denote the prophet benchmark and expected
online reward, respectively, with the model-specific expectations just given.

Throughout the paper, we ignore items that are not feasible even as a singleton, as well as those whose expected value is $0$. We use $(z)_+=\max\{z,0\}$ and $[a]=\{1,\ldots,a\}$ for positive integers $a$. Logarithms are natural unless a base is displayed.

\section{Lower Bound for the Standard Prophet Problem}

\begin{theorem}\label{thm:adv-lower}
For every sufficiently large $n$, there is a downward-closed instance on $n$
items with independent weighted Bernoulli values and a fixed arrival order,
known in advance, on which every online algorithm has competitive ratio
$\Omega(\log^2 n/\log\log n)$.
\end{theorem}

We keep the notation of the technical overview: layer $t$ has $d_t$ coordinates,
message $m_t\in\Sigma_M^{d_t}$, and codeword
$c_t=\enc_t(m_1,\ldots,m_{t-1})\in\Sigma_{C,t}^{d_t}$. Write
$m_{<t}=(m_1,\ldots,m_{t-1})$ and $m_{\le t}=(m_1,\ldots,m_t)$.
For an integer $v\ge1$, $\floorpowtwo{v}$ is the largest power of two not
exceeding $v$. The additional parameter $r$ is the number of physical items
in one group.
We first construct an instance for arbitrary integers $r,T\ge1$, and then
choose both parameters to fit $n$ items.

\subsection{The Construction}

Set
\begin{equation}\label{eq:adv-parameters}
 d_t=2^{t-1},\qquad p=\frac1{2rT},\qquad
 \Sigma_M=[(2rT)^{2r}],
\end{equation}
where $p$ is the activation probability of all items.
For each layer $t$, coordinate $i\in[d_t]$, message symbol $b\in\Sigma_M$, and
codeword symbol $s\in\Sigma_{C,t}$, create a group $G(t,i,b,s)$ of $r$
physical items. All these groups are disjoint. Every item of layer $t$
has value $1/(r d_t)$ with probability $p$, and zero otherwise.
A {\em message table} is a complete sequence of messages $(m_1, \dots, m_T)$ of length $T$, where each $m_t \in \Sigma_M^{d_t}$.
Each message table specifies one maximal feasible set:
\begin{equation}\label{eq:adv-feasible}
 \bigcup_{t=1}^T\ \bigcup_{i=1}^{d_t}
 G\bigl(t,i,(m_t)_i,(\enc_t(m_{<t}))_i\bigr).
\end{equation}
That is, in each layer $t$ and each coordinate $i$, we collect the group $G(t, i, (m_t)_i, (c_t)_i)$ where $c_t = \enc_t(m_{<t})$.
The feasibility system consists of all subsets of these sets. Thus at most
one group can be used at a coordinate, but all its members may be accepted.
The online arrival order is $T,T-1,\ldots,1$, with any fixed order within
each layer.

\paragraph{The code.}
One key component in the construction that remains unspecified is the code, captured by $\enc_t$.
Intuitively, if two tables first differ in layer $\tau$, we want their codewords to agree
on progressively fewer coordinates as $t-\tau$ grows.
The lemma below gives the existence of a code that is just strong enough for our purposes, while not requiring an overly large alphabet for codewords.
Such overly large alphabets would blow up the number of items and hurt the bound.
The proof of the lemma is largely a blackbox combinatorial argument.

\begin{lemma}[First-divergence-sensitive encoding]\label{lem:adv-code}
For the coordinate counts in \eqref{eq:adv-parameters}, there exist public
encoders $\enc_t$ with
\[
 \log|\Sigma_{C,t}|=O(t+\log|\Sigma_M|)
\]
such that, whenever two tables first differ in layer $\tau$, every $t>\tau$
satisfies
\begin{equation}\label{eq:adv-code-agreement}
 \frac{|c_t\cap c_t'|}{d_t}
 \le\frac1{4\floorpowtwo{t-\tau}^{\,2}}.
\end{equation}
Here $|c_t\cap c_t'|$ is the number of agreeing coordinates, as in the
overview. These encoders lead to at most
$n = \exp(O(T+r\log(2rT)))$ items per the construction.
\end{lemma}
\begin{proof}
We first give the elementary coding fact used in the construction. Suppose
there are $N$ possible inputs, the desired word length is $m$, and
$0<\delta\le1$. Put
\[
 k=\lceil\delta m\rceil,
 \qquad Q=\left\lceil\frac{2e}{\delta}N^{1/k}\right\rceil.
\]
A uniformly random word in $[Q]^m$ agrees with a fixed word on a prescribed
set of $k$ coordinates with probability $Q^{-k}$. Its probability of agreeing
on at least $k$ coordinates is therefore at most
\[
 \binom{m}{k}Q^{-k}
 \le\left(\frac{e}{\delta Q}\right)^k\le\frac{2^{-k}}N.
\]
Choose the words one at a time. Fewer than $N$ previously chosen words forbid
strictly less than the whole space of possible next words, so all $N$ inputs
can be assigned words with at most $k-1<\delta m$ agreements per pair.
The alphabet satisfies
\begin{equation}\label{eq:greedy-alphabet}
 \log Q=O\left(\log(2/\delta)+\frac{\log N}{\delta m}\right).
\end{equation}
This is an existence argument; the code is fixed before the item values are
drawn.

For each $h\in\{1,2,4,\ldots\}$ with $h<t$, apply this fact to all prefixes
$m_{\le t-h}$, using length $d_t$ and $\delta=1/(4h^2)$. At coordinate $i$,
let $(\enc_t(m_{<t}))_i$ be the tuple of the component symbols at that
coordinate. For $t=1$ the tuple is empty and the code alphabet has one symbol.
If the first divergence is $\tau<t$, take $h=\floorpowtwo{t-\tau}$. Since
$t-h\ge\tau$, the protected prefixes differ. Their component codewords
agree on fewer than $d_t/(4h^2)$ coordinates. Full tuples can agree only at
these coordinates, which proves \eqref{eq:adv-code-agreement}.

The number of message symbols in the prefix through $t-h$ is
\[
 \sum_{s=1}^{t-h}d_s=2^{t-h}-1<2d_t\,2^{-h}.
\]
Consequently \eqref{eq:greedy-alphabet} bounds the logarithm of the component
alphabet by
$O(\log(2h)+h^2 2^{-h}\log|\Sigma_M|)$.
Taking products of the alphabets, and hence sums of their logarithms, gives
\[
 \log|\Sigma_{C,t}|
 =O\left(\sum_{\substack{h<t\\h=1,2,4,\ldots}}
       \bigl(\log(2h)+h^2 2^{-h}\log|\Sigma_M|\bigr)\right)
 =O(t+\log|\Sigma_M|).
\]
For the last step, $\log(2h)\le O(h)$ and the dyadic $h<t$ have sum $O(t)$;
also $\sum_{h\ge1}h^2 2^{-h}<\infty$.
There are exactly $r|\Sigma_M|\sum_t d_t|\Sigma_{C,t}|$ physical items.
Using $\log|\Sigma_M|=2r\log(2rT)$ in the alphabet bound proves the claimed
item count.
\end{proof}

\paragraph{Remark.}
No code separates two choices of the current message immediately: when the
first divergence is in layer $\tau$, its inherited codewords in the same layer coincide. We
use the trivial overlap bound one in that layer. In later layers the bounds
in \eqref{eq:adv-code-agreement} sum to $O(1)$. This, rather than uniformly tiny overlap at every gap, is the property used below.

\subsection{The Prophet Benchmark}

\begin{lemma}\label{lem:adv-prophet}
For the construction in \eqref{eq:adv-parameters}--\eqref{eq:adv-feasible}, the expected prophet benchmark $\opt$ satisfies
\[
 (1-e^{-1})T\le\opt\le T.
\]
\end{lemma}
\begin{proof}
The upper bound follows because a feasible set contains at most $r d_t$
items of value at most $1/(r d_t)$ from each layer.
For the lower bound, build a message table in order $1,\ldots,T$, using only
the values of layers already visited.
Consider a partial prophet that sees all the realized values in each layer upon entering that layer.
In layer $t$, conditional on all choices and values
through layer $t-1$, the codeword $c_t$ is fixed. At each coordinate $i$,
there are $|\Sigma_M|$ groups with symbol $(c_t)_i$, indexed by the freely
chosen symbol $(m_t)_i$. Their values are still independent and have their
original laws. A group is entirely active with probability $p^r$. Hence the
probability that at least one of these groups is entirely active is
\[
 1-(1-p^r)^{|\Sigma_M|}
 \ge1-e^{-|\Sigma_M|p^r}
 =1-e^{-p^{-r}}\ge1-e^{-1}.
\]
Choose such a group when one exists, and any group otherwise. Choices at
different coordinates can be combined into one message $m_t$ without any
additional restriction. Each successful coordinate contributes $1/d_t$, so
the conditional expected contribution of layer $t$ is at least $1-e^{-1}$.
The union of the chosen groups is feasible by construction. Summing these
conditional expectations over layers proves the lower bound on $\opt$.
\end{proof}

\subsection{The Algorithm's Performance}

\paragraph{Intuition.}
The algorithm is forced to visit layers in the order of $T, \dots, 1$.
In each layer, the first acceptance at a coordinate can be unusually valuable: the algorithm
has inspected its value before choosing its group. After that acceptance,
unseen members of the chosen group have activation probability $p$, and the expected contribution is discounted by that probability.
We bound these two contributions separately. The first contributions are
counted ex post using two viable tables (this realizes the first-divergence intuition discussed in the technical overview); the remaining contributions are
bounded in expectation at the time the group is fixed (similar to the lower bound by \citet{BIK07}).

\begin{lemma}\label{lem:adv-algorithm}
Every online algorithm for the construction above, in the order
$T,T-1,\ldots,1$, has expected reward at most $2/r$.
\end{lemma}
\begin{proof}
Fix an optimal online algorithm and bound $\alg$.
Consider a realized execution of the algorithm.
At a certain time in the execution, a message table is \emph{viable} if its maximal feasible set contains every item accepted so far.
Feasibility ensures that some table is always viable.
We call a layer $t$ \emph{determined} if, at entry to layer $t$, all viable tables agree on $m_{\le t}$, and \emph{ambiguous} otherwise.
Further acceptances only remove viable tables.
Moreover, moving from layer $t$ to layer $t-1$ shortens the prefix on which agreement is required.
Thus determined layers form a suffix of the visited layers and ambiguous layers form a prefix.

Let $N_t$ be the number of ``touched'' coordinates of layer $t$, i.e., coordinates at which the algorithm accepts at least one item. We first prove
\begin{equation}\label{eq:adv-first-charge}
 \alg
 \le Tp+\frac1r\bE\left[
       \sum_{\substack{t:\text{ambiguous entry}}}\frac{N_t}{d_t}\right].
\end{equation}
At a determined entry to layer $t$, the common prefix fixes one group of items at every
coordinate of layer $t$. Realized values in this layer are still uncertain.
Conditional on the entry history, the sum of all values in these $r d_t$ items has expectation $p$, which is an upper bound on the algorithm's reward in that layer.

At an ambiguous entry to layer $t$, charge the first acceptance at each touched
coordinate the activated value of an item, $1/(r d_t)$. Immediately after that
acceptance, the group at the coordinate is fixed. There are $N_t$ touched coordinates, so the first acceptances contribute $N_t / (r d_t)$.
As for the subsequent acceptances in each group, conditional on the history right after the first acceptance, at most $r-1$ members of this group remain unseen; each still has expected value $p/(r d_t)$. Members seen and rejected earlier are unavailable.
All later acceptances at this coordinate therefore have conditional expected
total value at most $p/d_t$. Summing over the $d_t$ coordinates gives
at most $p$, which bounds the total value of the subsequent acceptances in the layer.
Finally, summing over layers $t$ gives \eqref{eq:adv-first-charge}.

It remains to bound the sum in \eqref{eq:adv-first-charge} on every realized play (i.e., we will bound the sum inside the expectation directly).
If no entry is ambiguous, it is zero.
Otherwise let $s$ be the last ambiguous layer in arrival order as perceived by the algorithm.
At its entry, choose any two viable tables that differ in their prefixes through $s$, and let $\tau\le s$ be the layer of their first divergence.\footnote{
    Note that $\tau$ may not be $s$, since acceptances made in layer $s$ may eliminate viable tables, and all we know is that {\em after} such eliminations, all viable tables agree up to layer $s - 1$.
}
Every layer $t>s$ has already finished. At every one
of its $N_t$ touched coordinates, both tables contain an accepted item.
Since groups are disjoint, both tables use the same group there; in particular, their codeword symbols agree.
Lemma~\ref{lem:adv-code} gives
\[
 \frac{N_t}{d_t}
 \le\frac1{4\floorpowtwo{t-\tau}^{\,2}}
 \le\frac1{4\floorpowtwo{t-s}^{\,2}}
 \qquad(t>s).
\]
The second inequality uses $\tau\le s$. Again, the pair of witness tables need not remain viable through layer $s$ as more acceptances are made, so we bound $N_s/d_s$ simply by one.
No layer $t<s$ has an ambiguous entry.
Consequently,
\[
 \sum_{\substack{t:\text{ambiguous entry}}}\frac{N_t}{d_t}
 \le1+\sum_{v=1}^{\infty}\frac1{4\floorpowtwo{v}^{\,2}}
 =1+\sum_{h=1,2,4,\ldots}\frac{h}{4h^2}=\frac32.
\]
The same pair of tables bounds all completed layers. Its retrospective
choice is legitimate because the argument is ex post.
Substitution into \eqref{eq:adv-first-charge}, with $Tp=1/(2r)$, proves the lemma.
\end{proof}

Now we are ready to prove Theorem~\ref{thm:adv-lower}.

\begin{proof}[Proof of Theorem~\ref{thm:adv-lower}]
The preceding two lemmas give ratio at least $(1-e^{-1})rT/2$.
For a sufficiently small absolute constant $c>0$, choose
\[
 r=\left\lfloor c\frac{\log n}{\log\log n}\right\rfloor,
 \qquad T=\lfloor c\log n\rfloor.
\]
Both are positive for sufficiently large $n$. Since
$\log(2rT)=O(\log\log n)$, the item bound in
Lemma~\ref{lem:adv-code} is at most $n$ after decreasing $c$ if necessary.
The resulting ratio is $\Omega(\log^2 n/\log\log n)$, as claimed.
\end{proof}

\section{Lower Bound for the Prophet Secretary Problem}

\begin{theorem}\label{thm:ps-lower}
For every sufficiently large $n$, there is a downward-closed instance on $n$
independent weighted Bernoulli items on which every prophet-secretary
algorithm has competitive ratio $\Omega(\log n)$, even when the entire
permutation is announced before play.
The same bound applies to the standard secretary problem by extension.
\end{theorem}

We continue to use $d_t$ for the coordinate count, $m_t$ for the message, and
$c_t=\enc_t(m_{<t})$ for the codeword in layer $t$. There is now one
item per pair of symbols at a coordinate, not a group. As discussed momentarily, the code in this construction encodes only the specified prefixes of earlier message symbols, which is different from the adversarial-order construction.

\subsection{The Construction}

Fix an integer $T\ge2$ and put
\begin{equation}\label{eq:ps-parameters}
 d_t=T2^{t-1},\qquad \Sigma_M=\{0,1\}^T,
 \qquad \Sigma_C=[2^{4T}].
\end{equation}
In particular, note that here we use the same codeword alphabet for all layers.
The subscript $t$ is therefore omitted.
For each $t\in[T]$, $i\in[d_t]$, $a\in\Sigma_M$, and $b\in\Sigma_C$, create
one item $(t,i,a,b)$. Its value realizes to $1/d_t$ with probability $2^{-T}$,
and zero otherwise, independently of every other item. Every message table
$(m_1,\ldots,m_T)$ specifies one maximal feasible set:
\[
 \{(t,i,(m_t)_i,(\enc_t(m_{<t}))_i):
            t\in[T],\ i\in[d_t]\}.
\]
Take all subsets of these sets are also feasible.
The number of  items is
\begin{equation}\label{eq:ps-size}
 2^{5T}\sum_{t=1}^T d_t
   =T(2^T-1)2^{5T}\le T2^{6T}.
\end{equation}
Each feasible set uses at most one item at a coordinate, so its value is at most one per layer and at most $T$ in total.

\paragraph{The code that encodes nested prefixes.}
In layer $t$, the encoder reads the first $t-\tau$ bits of every symbol of
$m_\tau$, for every layer $\tau<t$. Thus the layer-two codeword depends on one bit of every
layer-one message symbol.
The layer-three codeword depends on two bits of each layer-one message symbol
and one bit of each layer-two message symbol, etc.

The encoded prefixes are nested, so the codeword in a later layer encodes all the bits encoded by an earlier-layer codeword, as well as additional bits.
The doubling of $d_t$ ensures that the number of encoded bits is still proportional to the current codeword length.
Again, in the execution of an online algorithm, we call a message table viable if it is consistent with the items accepted so far.
The following lemma captures the guarantee we need from the code.

\begin{lemma}[Recovery of nested message prefixes]\label{lem:ps-code}
There are encoders with the alphabets in \eqref{eq:ps-parameters} such that
$c_t$ depends exactly on the protected bits as described, and any
\[
 k_t=d_t/T=2^{t-1}
\]
codeword symbols, together with their coordinate indices, determine those
bits uniquely. In particular, once a feasible algorithm has accepted $k_u$
items in layer $u$, all viable message tables have the same codeword $c_t$
for every $t\le u$, and the same first $u-t$ bits of each symbol of $m_t$.
At a coordinate in layer $t\le u$, at most $2^{T-u+t}$ items can
remain compatible with this recovered information.
\end{lemma}
\begin{proof}
Fix a public order in which to concatenate the encoded bits.
The total number of such bits in layer $t$ (i.e., those in the messages of earlier layers $\tau$ that are encoded by $c_t = \enc_t$) is
\[
 \sum_{\tau<t}(t-\tau)d_\tau
   =d_t\sum_{h=1}^{t-1}h2^{-h}<2d_t.
\]
The last strict inequality follows from
$\sum_{h\ge1}h2^{-h}=2$.
There are fewer than $2^{2d_t}$ possible strings of such encoded bits.
Assign each such string a codeword in $\Sigma_C^{d_t}$ greedily to minimize overlap with already chosen codewords.
For a fixed previous codeword, a uniformly random codeword agrees with it at any prescribed $k_t$ positions with probability $(2^{4T})^{-k_t}$.
The fraction of candidates agreeing in at least $k_t$ positions is at most
\[
 \binom{d_t}{k_t}2^{-4Tk_t}
   \le 2^{d_t-4d_t}=2^{-3d_t}.
\]
Even the union of the forbidden candidates from all fewer than $2^{2d_t}$
previous assignments has fraction below one. So a next codeword exists. Therefore
different strings of encoded bits receive codewords agreeing in fewer than
$k_t$ coordinates. In layer one, the encoded string is empty and its
single codeword can be chosen arbitrarily. Define $\enc_t$ by first reading
the encoded bits and then applying this fixed assignment.

Feasibility forces any $k_u$ accepted layer-$u$ items into distinct coordinates.
Their codeword symbols identify the encoded string uniquely by the
property of the code constructed and proved above. For $t<u$, the encoded string contains the first $u-t$ bits of each layer-$t$ message symbol. It also contains every bit
needed to compute $c_t$, because $u-\tau\ge t-\tau$ for every $\tau<t$.
The word $c_u$ itself is uniquely recovered as well. Consequently a
coordinate in layer $t\le u$ has a fixed codeword symbol and a fixed
message prefix of length $u-t$. There are exactly $2^{T-u+t}$ possible
completions of that prefix. At $t=u$ the prefix is empty; the codeword
symbol is fixed but all $2^T$ message symbols remain available. Every viable
table obeys these restrictions, which proves the last assertion.
\end{proof}

\paragraph{Remark (Reed-Solomon also suffices).}
The greedy construction is provided to be self-contained. One can
instead use the Reed-Solomon code \citep{ReedSolomon60} directly: pad the protected string to $2d_t$ bits, regard it as $k_t=d_t/T$
coefficients over $\mathbb F_{2^{2T}}$, and evaluate the resulting polynomial
of degree at most $k_t-1$ at $d_t$ distinct field points. Such points exist
because $d_t\le T2^{T-1}\le2^{2T}$. Any $k_t$ evaluations determine the
polynomial: the difference of two such polynomials has degree below $k_t$
and cannot have $k_t$ distinct roots unless it is zero. This gives the same
recovery property with an alphabet that embeds in $\Sigma_C$.

\subsection{The Prophet Benchmark}

\begin{lemma}\label{lem:ps-prophet}
The construction above satisfies
\[
 (1-e^{-1})T\le\opt\le T.
\]
Moreover, an online algorithm in the fixed layer order $1,\ldots,T$ achieves
expected reward at least $(1-e^{-1})T$.
\end{lemma}
\begin{proof}
The upper bound is immediate from at most one item per coordinate and value
$1/d_t$ per item. For the lower bound, it suffices to describe the stated
online algorithm. Before processing layer $t$, it has completed messages
$m_1,\ldots,m_{t-1}$, so it can compute $c_t$. At each coordinate, it accepts
the first active item whose codeword symbol agrees with $c_t$ and ignores
all other items at that coordinate. At the end of the layer, fill any
message coordinates at which no item was accepted with arbitrary symbols.
Every such completion is allowed and is consistent with all previous
acceptances. Thus the algorithm remains feasible.

Conditional on the earlier layers, a coordinate offers $2^T$ independent
items with the codeword symbol fixed. Each is active with probability
$2^{-T}$, so the probability of at least one success is
\[
 1-(1-2^{-T})^{2^T}\ge1-e^{-1}.
\]
A successful coordinate contributes $1/d_t$. Summing first over coordinates
and then over layers proves the expected-reward guarantee. The prophet can
choose at least the same feasible set on every realization.
\end{proof}

\subsection{Constrained and Unconstrained Acceptances}

In order to bound the performance of online algorithms, we need to define a few auxiliary concepts.
Immediately before each item's value is revealed, define the \emph{cutoff
layer} by
\[
 u=\max\bigl(\{t:\text{at least }k_t\text{ items have already been accepted
 in layer }t\}\cup\{0\}\bigr),
 \qquad k_t=d_t/T.
\]
The cutoff never decreases throughout the algorithm's execution.
An acceptance in layer $t>u$ is \emph{unconstrained}, and one in layer $t\le u$ is \emph{constrained} ($t$ is the layer containing the item being accepted, and $u$ is the cutoff layer right before that acceptance).
Note that these names refer only to the restrictions certified by the cutoff; other
accepted items can impose further feasibility restrictions.
The state used for the above classification is always the one before the current value is observed.
In particular, an acceptance that raises $u$ is unconstrained.

Below we first bound the total contribution of unconstrained acceptances, which is the easier part of the argument.

\begin{lemma}\label{lem:unconstrained}
The total realized reward of unconstrained acceptances is at most one.
\end{lemma}
\begin{proof}
Once the $k_t$-th item is accepted in layer $t$, the cutoff is at least $t$
forever. Hence layer $t$ can contribute at most $k_t$ unconstrained
acceptances, including the acceptance that first reaches the threshold.
Every item in the layer is worth at most $1/d_t$, so their total value is at
most
\[
 \sum_{t=1}^T\frac{k_t}{d_t}=1.
\]
This argument is ex post and makes no restriction on how observed values
influence the acceptance decisions.
\end{proof}

It remains to bound the expected reward of constrained acceptances. At a
positive cutoff $u$, Lemma~\ref{lem:ps-code} gives a uniquely determined
codeword $c_u$ and restricts the eligible identities in all layers $t\le u$.
For a fixed codeword $c_u$, the total mean of all compatible items in
layer $t$ is at most $2^{t-u}$, so the sum over these layers is less than two.
This is a mean calculation that gives an intuitive bound, not yet a formal one that works for the actual algorithm.
The next reduction addresses that distinction.

\subsection{The Harvesting Game}

Consider the following relaxed setting that grants more power to the algorithm (or the {\em harvester} in this setting).
It focuses solely on constrained acceptances; unconstrained acceptances have already been bounded earlier.
Generate the permutation of items by assigning each item independently to a
uniform bin in $[T]$, ordering each bin uniformly at random, and concatenating the
bins. The resulting permutation is uniform. Bin sizes are random. Announce the entire permutation and the bin assignments before the first arrival; write $\mathcal D$ for this announcement.
Note that all this is independent of all item values.

A harvester keeps a cutoff $u\in\{0,\ldots,T\}$ and, when $u>0$, a declaration
$c_u$ that is a possible codeword in layer $u$. The declaration determines
codeword symbols and message prefixes in layers $t\le u$ as in
Lemma~\ref{lem:ps-code}. All items in these layers whose message and codeword symbols are consistent with $c_u$ are considered {\em compatible}.
Initially $u=0$, $c_u$ is empty, and there are no compatible items.
The rules at the beginning of a bin are as follows. First the harvester
collects the entire value of every compatible item in that bin under its
current declaration, inherited from the previous bin. It may then make any number of strict cutoff increases, where it replaces $u$ with an integer between $u + 1$ and $T$.
At each increase, it declares an arbitrary possible codeword $c_u$ for
the new layer $u$ and collects every item of the current bin compatible with the newly declared $u$ and $c_u$.
Successive declarations need not be consistent. We ignore feasibility
restrictions (which gives the harvester more power), but an item contributes only once, at its first harvest.
Crucially, the declaration is chosen before the newly harvested values are revealed (otherwise the harvester would be too powerful).
The harvester observes no other values.
It may use earlier harvests to decide on further increases and declarations, and it carries its last declaration into the next bin.
There are at most $T$ cutoff increases, in addition to one carried harvest
per bin. Several increases in the same bin are permitted and can collect
different items.

The following lemma connects the harvesting game to constrained acceptances made by an actual online algorithm.

\begin{lemma}[Reduction to harvesting]\label{lem:harvesting-reduction}
Conditional on any announcement $\mathcal D$, the expected reward of the
constrained acceptances of any online algorithm (that can even depend on $\mathcal{D}$) is at most the optimal
conditional expected reward in the harvesting game.
\end{lemma}
\begin{proof}
Fix $\mathcal D$.
Given an online algorithm, we construct a strategy in the harvesting game that produces at least as much conditional expected reward as the algorithm's constrained acceptances.
For each item $i$, let $J_i$ be the indicator that, immediately before its value is
revealed, its layer satisfies $t\le u$ and its identity is compatible with
the codeword $c_u$ already determined by previous acceptances. This is an
eligibility indicator, not an acceptance indicator. Every constrained
acceptance has $J_i=1$, by Lemma~\ref{lem:ps-code}, and nonnegative values
therefore give the ex post upper bound
\[
 \text{constrained reward}\le\sum_iJ_iX_i.
\]
Given $\mathcal D$ and the history immediately before this arrival, $J_i$ is
known and $X_i$ retains its prior mean. Independence gives
\begin{equation}\label{eq:predictable-eligibility}
 \bE[J_iX_i\mid\mathcal D]
     =\bE[J_i\mid\mathcal D]\,\bE[X_i].
\end{equation}
This holds even though the subsequent acceptance decision can depend on
$X_i$. If that decision raises the cutoff, its own reward is already
unconstrained; the new codeword is used only for subsequent eligibility.

Fix a realized execution of the original algorithm. In each bin, record its
carried cutoff upon entry and every subsequent strict increase, with the corresponding
uniquely determined codewords. Move that entire list of declarations to
the opening of the bin and keep their order.
This is a structurally legal schedule (not to be confused with an actual strategy; see discussion below): cutoffs increase strictly when changed, and arbitrary declarations are allowed. While the cutoff is fixed, the recovered codeword cannot change, because further acceptances only remove viable tables.
The problem is that the schedule depends on the realized values, and therefore cannot be implemented in the stated form by the harvester who does not see the values upon entering a bin.
We will turn the schedule into an actual strategy later through simulation of the algorithm on phantom values.

Back to the above schedule: every item with $J_i=1$ lies in a harvest of this schedule, because its pre-arrival declaration occurs in the list for its own bin. Therefore
\[
 \sum_iJ_i\bE[X_i]
 \le \sum_{i \text{ is harvested}}\bE[X_i].
\]
Here, the right side sums over the union of the recorded harvests, counting each item only once.

To implement an actual, executable strategy, before any arrival, draw independent phantom
values from all the priors and independent copies of the original algorithm's
random coins. Simulate that algorithm on these virtual values with the same
announcement $\mathcal D$, and record the resulting schedule as above.
Execute this schedule on the actual values, without using the actual harvest
observations to change its declarations. This is legal: all declarations
are chosen using only $\mathcal D$ and private randomness, independently of
the actual values. Conditional on the simulated schedule, its expected
harvest is exactly the sum of the prior means in its harvested union.

The simulated execution has the same conditional distribution as the original
execution. Consequently the expected sum of means in its recorded union is
at least $\bE[\sum_iJ_i\bE[X_i]\mid\mathcal D]$, by the ex post inequality
above. Applying \eqref{eq:predictable-eligibility} and summing over items gives
\begin{align*}
 \bE[\text{harvesting reward}\mid\mathcal D]
 &\ge \bE\left[\sum_iJ_i\bE[X_i]\,\middle|\,\mathcal D\right]\\
 &=\bE\left[\sum_iJ_iX_i\,\middle|\,\mathcal D\right]\\
 &\ge\bE[\text{constrained reward}\mid\mathcal D].
\end{align*}
The $J_i$ and $X_i$ in the last two expressions refer to the original online
execution; equality in distribution justifies the comparison with the phantom
record.
This proves the reduction lemma.
\end{proof}

\subsection{Bounding the Contribution of Constrained Acceptances}

With the reduction to the harvesting game established, we are ready to bound the contribution of constrained acceptances.
For a possible declaration $c_u$, let its \emph{menu} $M(c_u)$ be all items
in layers $t\le u$ compatible with the codeword symbols and message prefixes
recovered from $c_u$. Put $M(c_0)=\varnothing$. A harvest in bin $B_j$ collects
$M(c_u)\cap B_j$.

As in the overview, a \emph{piece} $P_{t,i,s,p}$ fixes a layer $t$, a
coordinate $i$, a codeword symbol $s\in\Sigma_C$, and a binary prefix $p$ of
length at most $T$. It consists of all items $(t,i,a,s)$ with $a$ extending $p$;
hence $|P_{t,i,s,p}|=2^{T-|p|}$, where $|p|$ is the length of $p$ as a binary string. Every menu is a disjoint union of one piece
per coordinate of each layer $t\le u$. In that layer the prescribed prefix
has length $u-t$. Its piece therefore has size $2^{T-u+t}$, and each of its
items has mean $1/(2^T d_t)$.

\paragraph{Intuition: all menus are evenly distributed across bins.}
A fixed menu has total mean less than two. A uniform bin should receive
about a $1/T$ fraction of this mean. The algorithm knows the bin assignment,
so this estimate must hold simultaneously for all possible declarations of codewords.
There are too many complete declarations for a direct union bound. Instead,
there are fewer than twice as many pieces as items.
Concentrating each piece suffices for every menu.

\begin{lemma}[Uniform bin bounds for pieces and menus]\label{lem:pieces-bins}
With probability at least
\[
 1-2T^2\exp\bigl(-(10-6\log2)T\bigr)
\]
over the independent bin assignments, every piece $P_{t,i,s,p}$ and bin
$B_j$ satisfy
\begin{equation}\label{eq:piece-occupancy}
 |P_{t,i,s,p}\cap B_j|
 \le\frac{e-1}{T}|P_{t,i,s,p}|+10T.
\end{equation}
On the same event, for every possible declaration $c_u$ and every bin,
\begin{equation}\label{eq:menu-mean}
 \sum_{a\in M(c_u)\cap B_j}\bE[X_a]
 \le\frac{2(e-1)}T+\frac{10T^2}{2^T}.
\end{equation}
The second bound concerns the sum of the prior means, conditional on the announced (indices of) items and bins.
\end{lemma}
\begin{proof}
Fix a piece $P$ and bin $B_j$. The independent bin assignments make
$|P\cap B_j|$ binomial with parameters $|P|$ and $1/T$. Therefore
\[
 \bE[e^{|P\cap B_j|}]
  =\bigl(1+(e-1)/T\bigr)^{|P|}
  \le \exp\bigl((e-1)|P|/T\bigr).
\]
Markov's inequality gives
\[
 \Pr\bigl[|P\cap B_j|>(e-1)|P|/T+10T\bigr]\le e^{-10T}.
\]
For each coordinate and codeword symbol there are $2^{T+1}-1$ prefixes,
including the empty prefix and all full strings. Hence the number of pieces
is less than twice the number of physical items in \eqref{eq:ps-size},
and the number of piece--bin pairs is at most $2T^2 2^{6T}$.
The union bound proves \eqref{eq:piece-occupancy} simultaneously, with the
stated failure probability. This event depends only on the bin assignments.

Fix any assignment on this event, any declaration, and any bin. Summing
\eqref{eq:piece-occupancy} over the $d_t$ pieces of each layer in the menu,
and weighting each item by its mean, yields
\begin{align*}
 \sum_{a\in M(c_u)\cap B_j}\bE[X_a]
 &\le\sum_{t=1}^{u}\frac{d_t}{2^T d_t}
      \left(\frac{e-1}{T}2^{T-u+t}+10T\right)\\
 &\le\frac{e-1}{T}\sum_{t=1}^{u}2^{t-u}
             +\frac{10uT}{2^T}
 \le\frac{2(e-1)}T+\frac{10T^2}{2^T}.
\end{align*}
For $u=0$ the menu is empty. This proves the uniform menu bound.
\end{proof}

\paragraph{Remark.}
The additive term in \eqref{eq:piece-occupancy} cannot in general be removed
for each piece: a singleton has occupancy one in its assigned bin, however
small its mean occupancy $1/T$ was. Its aggregated mean-value cost is
negligible because each item's activation probability includes $2^{-T}$.

Now we can bound the contribution of constrained acceptances.

\begin{lemma}\label{lem:constrained-reward}
Every online algorithm, even given $\mathcal D$, has expected
reward from constrained acceptances at most
\[
 4(e-1)+\frac{20T^3}{2^T}
   +2T^3\exp\bigl(-(10-6\log2)T\bigr)=4(e-1)+o(1).
\]
\end{lemma}
\begin{proof}
Fix an announcement $\mathcal{D}$ satisfying Lemma~\ref{lem:pieces-bins}.
Any execution of any harvesting strategy opens at most $T$ carried menus and $T$ new menus from strict cutoff increases.
Immediately before each harvest, condition on the harvester's observations and random choices so far. Every newly harvested item is still unobserved and contributes its mean. Thus the conditional expected gain from this harvest is at most the sum of the means in the current bin--menu intersection. Summing these conditional bounds over at most $2T$ harvests bounds the strategy's expected reward.
By \eqref{eq:menu-mean}, this is at most
\[
 2T\left(\frac{2(e-1)}T+\frac{10T^2}{2^T}\right)
       =4(e-1)+\frac{20T^3}{2^T}.
\]
The bound applies to all strategies selected using the announcement.
Lemma~\ref{lem:harvesting-reduction} therefore gives the same upper bound
for the conditional expected constrained reward of the original algorithm.

On an exceptional announcement (where concentration fails), the original algorithm still returns a feasible set of realized value at most $T$.
Multiplying this bound by the failure probability in Lemma~\ref{lem:pieces-bins} and averaging proves the claim.
\end{proof}

This finally gives us all the pieces needed to prove the main theorem in this section.

\begin{proof}[Proof of Theorem~\ref{thm:ps-lower}]
Lemmas~\ref{lem:unconstrained} and \ref{lem:constrained-reward} bound the
online value by $1+4(e-1)+o(1)$, while Lemma~\ref{lem:ps-prophet} gives
$\opt\ge(1-e^{-1})T$. For a small enough absolute constant $c>0$, choose
$T=\lfloor c\log n\rfloor$. Equation~\eqref{eq:ps-size} then gives at most
$n$ genuine items.
This gives the desired lower bound of $\Omega(T)=\Omega(\log n)$ for the prophet secretary problem.
The same bound holds for the standard secretary problem.
\end{proof}

\section{Upper Bound for the Standard Prophet Problem}

\begin{theorem}\label{thm:adv-upper}
For independent nonnegative integrable values and arbitrary downward-closed
feasibility on $n$ items, the standard prophet problem admits competitive
ratio $O(\log^2 n/\log\log n)$.
\end{theorem}

For a residual downward-closed system $\cS$ on the unseen items, let
$\OPT(\cS)$ denote its realized prophet benchmark and $\alg(\cS)$ the optimal
expected additional online reward in the inherited order. Neither includes
rewards already obtained. For a next item $i$, use the deletion and
contraction notation from the overview:
\[
 \cS\setminus i=\{S\in\cS:i\notin S\},\qquad
 \cS/i=\{S\setminus\{i\}:S\in\cS,\ i\in S\}.
\]
Downward closure makes contraction a subset of deletion.
These definitions are used in the continuation
arguments in this section and the next one.

The core of the proof is an improved bound for the case with binary values.
This immediately extends to the general problem through a blackbox reduction by \citet{Rub16}.
We first present the bound for binary values and show how it implies the main theorem of the section.

\begin{lemma}[Binary prophet bound]\label{lem:binary-bound}
If all item values lie in $\{0,1\}$, then in every residual instance defined by $\cS$ with
$m$ items, and for every $a\ge2$,
\begin{equation}\label{eq:binary-invariant}
 \bE\bigl[a^{\OPT(\cS)}\bigr]
 \le\exp\bigl(2(a+\log(m+1))\alg(\cS)\bigr).
\end{equation}
As a result, the binary standard prophet problem has competitive ratio
$O(\log n/\log\log n)$. Both statements also hold when the arrival-order
adversary may choose the next unseen identity using only previously revealed
values and decisions, and $\alg$ is the optimal guaranteed reward against
that adversary.
\end{lemma}

\begin{proof}[Proof of Theorem~\ref{thm:adv-upper}]
The value-band extension of \citet[proof of Corollary~1]{Rub16}
turns a binary prophet guarantee into a general-value guarantee at an
additional factor $O(\log(\rho+1))$, where $\rho$ is the largest cardinality
of a feasible set. Apply this black-box extension with the binary guarantee
in Lemma~\ref{lem:binary-bound}. Since $\rho\le n$, the resulting ratio is
\[
 O\left(\frac{\log n}{\log\log n}\log(\rho+1)\right)
       =O\left(\frac{\log^2 n}{\log\log n}\right).\qedhere
\]
\end{proof}

Now we come back to the binary case.
The proof analyzes the optimal online algorithm, which is implicit but satisfies certain recurrences that follow from optimality.
The ideas are discussed below.

\paragraph{Intuition for the binary induction.}
From any residual problem onward, fixing the next item to arrive, there are two possible continuations: either we accept the item, which gives immediate reward but limits future options, or we reject the item and keep more freedom.
The two continuations apply to both the prophet and the algorithm.
The prophet can compare both feasible continuations after seeing the future,
whereas the algorithm compares only their expectations.
Examining the moment generating function of the prophet benchmark lets us charge the prophet's extra power to the online gain up to a factor of $O(\log n / \log \log n)$.
If accepting an active item gives the algorithm a large improvement, that improvement pays for the extra power directly.
Otherwise, acceptance has already lost at least half a unit of expected continuation value.
The exponential penalty for that loss pays for the extra power instead.

\begin{proof}[Proof of Lemma~\ref{lem:binary-bound}]
Fix $a\ge2$ and put $b_m=2(a+\log(m+1))$. We prove
\eqref{eq:binary-invariant} by induction on $m$. With no remaining items,
both $\OPT$ and $\alg$ are zero and the two sides equal one.
Let $i$ be the next item. Abbreviate the continuation values by
$\alg_0=\alg(\cS\setminus i)$ and $\alg_1=\alg(\cS/i)$, and the
corresponding offline values by $\OPT_0,\OPT_1$ on the same future realization.
If $i$ is a loop (i.e., it is infeasible even as a singleton), then the contraction continuation does not exist, and we use the induction hypothesis for deletion and
$b_m\ge b_{m-1}$.
Now assume $i$ is feasible.
Downward closure gives $\alg_0\ge\alg_1$ and $\OPT_0\ge \OPT_1$.
Independence leaves the remaining priors unchanged after observing $X_i$.
Writing $p_i=\Pr(X_i=1)$ and
$\Delta=(1+\alg_1-\alg_0)_+\in[0,1]$, the online recurrence is
\begin{equation}\label{eq:binary-bellman}
 \alg=\alg_0+p_i\Delta.
\end{equation}
An inactive item may be rejected without loss.

The two offline continuation values are integers. If $X_i=1$, including
$i$ improves the deletion optimum precisely when $\OPT_1=\OPT_0$, and then by
one. If $X_i=0$, the optimum equals $\OPT_0$. Therefore
\begin{align}
 \bE[a^{\OPT(\cS)}]
 &=\bE[a^{\OPT_0}]
   +p_i(a-1)\bE\bigl[a^{\OPT_1}\ind_{\{\OPT_1=\OPT_0\}}\bigr]\notag\\
 &\le\bE[a^{\OPT_0}]+p_i(a-1)\bE[a^{\OPT_1}].
 \label{eq:binary-moment-step}
\end{align}
We compare the inductive bounds for these two terms with $e^{b_m\alg}$.

\emph{Case 1: $\Delta>1/2$.}
In this case $p_i\le2p_i\Delta$. Both child moments are at most
$e^{b_m\alg_0}$, since $b_{m-1}\le b_m$ and $\alg_1\le\alg_0$.
Consequently
\begin{align*}
 \bE[a^{\OPT(\cS)}]
 &\le e^{b_m\alg_0}\bigl(1+2(a-1)p_i\Delta\bigr)\\
 &\le e^{b_m\alg_0+2(a-1)p_i\Delta}
 \le e^{b_m(\alg_0+p_i\Delta)}=e^{b_m\alg}.
\end{align*}
Here we used $1+x\le e^x$, $b_m\ge2(a-1)$, and
\eqref{eq:binary-bellman}.

\emph{Case 2: $\Delta\le1/2$.}
Now $\alg_0-\alg_1\ge1/2$, so $\alg_0\ge1/2$ and
$\alg-\alg_1\ge1/2$. For the deletion term, induction gives
\[
 \frac{\bE[a^{\OPT_0}]}{e^{b_m\alg}}
 \le e^{-(b_m-b_{m-1})\alg_0}
 \le e^{-(b_m-b_{m-1})/2}=\frac{m}{m+1}.
\]
For the contraction term, using $p_i\le1$ and $\alg_1\ge0$ gives
\[
 \frac{p_i(a-1)\bE[a^{\OPT_1}]}{e^{b_m\alg}}
 \le(a-1)e^{-b_m(\alg-\alg_1)}
 \le(a-1)e^{-b_m/2}
 =\frac{(a-1)e^{-a}}{m+1}\le\frac1{m+1}.
\]
The two bounds sum to at most one, proving the induction step.

For an adversary that sees what the algorithm sees and controls the remaining part of the arrival order adaptively, define each residual $\alg(\cS)$ by
backward induction in the finite-horizon zero-sum game. The adversary first chooses
an unseen item; after that item's value is revealed, the algorithm
chooses the better of deletion and contraction. Thus $\alg(\cS)$ is the
minimum over (indices of) items of the right side of \eqref{eq:binary-bellman},
with the corresponding guaranteed continuation values. Choose a minimizing
item $i$ for the induction above. All its steps still hold: the unseen
values remain independent with the original distributions, and the 
deletion continuation has value at least that of contraction. This proves the
same invariant for the adversarial game.

Finally, Jensen's inequality yields
\[
 (\log a)\opt\le\log\bE[a^{\OPT}]\le b_n\alg.
\]
Taking $a=\max\{2,\log(n+1)\}$ makes $b_n=O(\log n)$ and
$\log a=\Theta(\log\log n)$ as $n$ grows. The resulting ratio is
$O(\log n/\log\log n)$, completing the proof.
\end{proof}

\section{Upper Bound for the Free-Order Prophet Problem}

\begin{theorem}\label{thm:free-upper}
Let $k$ be the least integer at least two such that $k^k\ge2n+2$.
For every downward-closed instance with independent nonnegative integrable
values, there is an order $\sigma$, chosen from the priors, for
which
\[
 \opt(X)\le(25k+2)\alg_\sigma(X).
\]
In particular, the free-order prophet problem has competitive ratio
$O(\log n/\log\log n)$.
\end{theorem}

Throughout this section the feasibility family is fixed; we write
$\opt(W)$ and $\alg_\sigma(W)$ to indicate which random values $W$ / distributions thereof are used.
For a nonnegative variable $W$, its \emph{excess function} is
\[
 \excess_W(v)=\bE[(W-v)_+]\qquad(v\ge0).
\]
It satisfies the following basic scaling identity: $\excess_{aW}(v)=a\excess_W(v/a)$ for $a>0$.
If $Y$ is a mixture with sampling probabilities $w_j/W$ (where $W = \sum_j w_j$) and conditional distributions $Y_j$, then
$\excess_Y(v)=\sum_j(w_j/W)\excess_{Y_j}(v)$.
We use the deletion and contraction notation from the preceding section inside continuation arguments.

\subsection{Constant Bound for Activated Exponentials}

An activated exponential $Y$ of scale $s>0$ equals an exponential variable of
mean $s$ with probability $p\in[0,1]$, and zero otherwise, with independent
activation.
Its excess function is
\[
 \excess_Y(v)=ps e^{-v/s}.
\]
This can be obtained by integrating the tail probability $\Pr(Y>z)=p e^{-z/s}$ over $z\ge v$.
The zero variable is allowed by taking $p=0$ and any positive scale.
Below we will prove a result for activated exponentials, that essentially says the free-order prophet problem admits a constant ratio restricted to activated exponential values.
This motivates the reduction of the general problem to activated exponentials, which we discuss later.
We use a similar comparison between recurrences as in the proof of the earlier binary-value upper bound.

\paragraph{Intuition.}
For a feasible next item of value $W$, let $\alg_0$ and $\alg_1$ be the
optimal future rewards after rejecting or accepting it. Their difference
is the opportunity cost of acceptance, and is nonnegative by downward
closure. The online value is exactly
\begin{equation}\label{eq:excess-bellman}
 \bE[\max\{\alg_0,W+\alg_1\}]
   =\alg_0+\excess_W(\alg_0-\alg_1).
\end{equation}
For an activated exponential, this increment has the same exponential
shape that appears when comparing the two offline continuation moments.
Sorting by scale ensures that both continuation moments can be compared at the
current scale.

\begin{lemma}[Decreasing-scale order]\label{lem:exponential-order}
Let $Y_1,\ldots,Y_n$ be independent activated exponentials with scales
$s_1,\ldots,s_n$. If $\sigma$ lists the items in nonincreasing order of
$s_i$, then $\opt(Y)\le2\alg_\sigma(Y)$.
\end{lemma}
\begin{proof}
In a residual instance, let $s$ be its largest remaining scale, $\OPT$ its
realized prophet benchmark, and $\alg$ its optimal online continuation
value in the inherited order. We prove by induction that
\begin{equation}\label{eq:exp-invariant}
 \bE[e^{\OPT/(2s)}]\le e^{\alg/s}.
\end{equation}
An empty continuation has moment one. If a nonempty continuation has largest
scale $s'\le s$, then its inductive bound is valid also at denominator
$2s$. Specifically, for its benchmark $\OPT_j$ and online value $\alg_j$,
concavity of $z\mapsto z^{s'/s}$ gives
\[
 \bE[e^{\OPT_j/(2s)}]
 \le\bigl(\bE[e^{\OPT_j/(2s')}]\bigr)^{s'/s}
 \le e^{\alg_j/s}.
\]
An empty continuation satisfies the same inequality. If the current item is infeasible, only the deletion continuation is needed.

Otherwise let $Y$ be the current reward, whose scale is $s$, and let $p$
be its activation probability. On the same realization of the future
values, downward closure gives $\OPT_1\le \OPT_0$ and
$\OPT=\max\{\OPT_0,Y+\OPT_1\}$. Thus
\[
 e^{\OPT/(2s)}
 \le e^{\OPT_0/(2s)}+(e^{Y/(2s)}-1)e^{\OPT_1/(2s)}.
\]
To verify this inequality, the added term is nonnegative if the deletion
branch wins; if the acceptance branch wins, subtract its value and use
$e^{\OPT_0/(2s)}\ge e^{\OPT_1/(2s)}$.
An exponential of mean $s$ has moment
\[
 \int_0^\infty e^{z/(2s)}e^{-z/s}\,\frac{dz}{s}=2
\]
at $1/(2s)$. Consequently $\bE[e^{Y/(2s)}-1]=p$. The current reward is independent
of both continuation benchmarks, so the continuation bounds imply
\begin{align*}
 \bE[e^{\OPT/(2s)}]
 &\le e^{\alg_0/s}+p e^{\alg_1/s}\\
 &=e^{\alg_0/s}\bigl(1+p e^{-(\alg_0-\alg_1)/s}\bigr)\\
 &\le\exp\left(\frac{\alg_0+ps e^{-(\alg_0-\alg_1)/s}}s\right)
 =e^{\alg/s}.
\end{align*}
The last identity is the online recurrence \eqref{eq:excess-bellman}, with $\excess_Y(v)=ps e^{-v/s}$.
This proves \eqref{eq:exp-invariant}.
At the initial state $\cF$, Jensen's inequality gives
$\opt(Y)/(2s)\le\log\bE[e^{\OPT/(2s)}]\le\alg_\sigma(Y)/s$.
Multiplying by $2s$ proves the lemma.
\end{proof}

\subsection{The Reduction via Excess Domination}

We seek a short list of activated exponential distributions $Y_{i,j}$ for each item,
and a sampling distribution $\Daux_i$ on their indices. After sampling
$j_i$ independently from each $\Daux_i$, draw $Y_i$ from the conditional distribution
$Y_{i,j_i}$. All these choices and draws are independent across items. The
algorithm observes the original values $X_i$, not the auxiliary draws; the
auxiliaries are used merely to determine an order and serve in its analysis.

To state exactly which comparisons are required, for each item $i$, write
\[
 \mu_i=\bE[X_i]>0,\qquad U_i=\min\{X_i/\mu_i,2n\}.
\]
The auxiliary distributions are in these normalized units.
The normalization $U_i=\min\{X_i/\mu_i,2n\}$ ensures both $U_i\le2n$ and
$\bE[U_i]\le1$.
The capping is without loss of generality, since very large values can be collected by an algorithm taking the first item above its cap.
Its cost is controlled in Lemma~\ref{lem:normalization-cost} below.
The desired relations, for every $v\ge0$, are
\begin{align}
 \excess_{Y_{i,j}}(v)&\le\excess_{U_i}(v)+\frac1n
                &&\text{for every }j,\label{eq:conditional-domination}\\
 \excess_{(U_i-k)_+}(v)&\le\excess_{6kY_i}(v)
                &&\text{in expectation over }j_i \sim \Daux_i.
                \label{eq:mixture-domination}
\end{align}
Here, the cost from shifting $U_i$ by $k$ is covered by preselecting a fixed feasible set of large total mean.
The following lemma formally controls the costs of capping and shifting.

\begin{lemma}[Cost of capping and shifting]\label{lem:normalization-cost}
Let $\mu_i=\bE[X_i]>0$ and $U_i=\min\{X_i/\mu_i,2n\}$ each item $i$.
For every fixed order $\sigma$,
\begin{itemize}
    \item The contribution from realized values above $2n$ is controlled:
    \begin{equation}
     \sum_i\bE[(X_i-2n\mu_i)_+]\le2\alg_\sigma(X),
                      \label{eq:cap-cost}
    \end{equation}
    \item The offset by $1/n$ in \Cref{eq:conditional-domination} is controlled:
    \begin{equation}
     \max_{S\in\cF}\sum_{i\in S}\mu_i\le\alg_\sigma(X),
     \qquad \frac1n\sum_i\mu_i\le\alg_\sigma(X),
                  \label{eq:mean-cost}
    \end{equation}
    \item The cost of shifting $U_i$ by $k$ is controlled:
    \begin{equation}
     \opt(X)\le(2+k)\alg_\sigma(X)
                +\opt\bigl((\mu_i(U_i-k)_+)_i\bigr).
                    \label{eq:total-reduction-cost}
    \end{equation}
\end{itemize}
\end{lemma}
\begin{proof}
Markov's inequality gives $\Pr(X_i>2n\mu_i)\le1/(2n)$. Consider the algorithm
that takes the first item exceeding its cap, and no other item. Every
singleton under consideration is feasible. The probability that no item
preceding $i$ exceeds its cap is at least $1/2$, by the union bound; this
event is independent of $X_i$. On an exceedance the policy earns at least
$(X_i-2n\mu_i)_+$. Summing over the possible accepted identities gives
expected reward at least half the left side of \eqref{eq:cap-cost}.
The optimal online algorithm is at least as good.

An algorithm that preselects a feasible set maximizing the sum of its means
proves the first bound in \eqref{eq:mean-cost}. An algorithm that preselects the
single item of largest mean proves the second, since there are at most $n$
retained items and $\sum_i\mu_i/n\le\max_i\mu_i$.

On every realization, capping can reduce the value of a feasible set by
at most $\sum_i(X_i-2n\mu_i)_+$. Also
\[
 \mu_iU_i\le k\mu_i+\mu_i(U_i-k)_+.
\]
Apply this inequality on each feasible set and maximize. Taking
expectations, \eqref{eq:cap-cost} and \eqref{eq:mean-cost} yield
\eqref{eq:total-reduction-cost}. If no retained item remains, all three
assertions are immediate with zero benchmark and reward.
\end{proof}

The next lemma shows that a construction satisfying these conditions implies the main theorem of this section.
We will discuss the actual construction later.

\begin{lemma}[Reduction through excess domination]\label{lem:excess-reduction}
Suppose that, for every item $i$, an independent auxiliary family of activated exponentials $\{Y_{i, j}\}$ and sampling distribution $\Daux_i$ satisfy
\eqref{eq:conditional-domination}--\eqref{eq:mixture-domination}.
Let $s_{i,j}$ be the scale of $Y_{i,j}$.
Independently sample the indices, sort the original items in
nonincreasing order of $\mu_i s_{i,j_i}$, and run the optimal online
algorithm in that order. This policy satisfies
\[
 \opt(X)\le(25k+2)\bE_\sigma[\alg_\sigma(X)].
\]
Therefore some fixed order attains the ratio in
Theorem~\ref{thm:free-upper}.
\end{lemma}
\begin{proof}
We use the bounds of Lemma~\ref{lem:normalization-cost} below to handle the loss from capping, shifting, and offsetting.

\emph{Conditional online comparison.}
Fix all sampled indices. After scaling back by $\mu_i$, the auxiliary rewards
$\mu_iY_{i,j_i}$ are independent activated exponentials, and $\sigma$ is
their decreasing-scale order. Scaling
\eqref{eq:conditional-domination} gives
\[
 \excess_{\mu_iY_{i,j_i}}(v)
 \le\excess_{\mu_iU_i}(v)+\frac{\mu_i}{n}
 \le\excess_{X_i}(v)+\frac{\mu_i}{n}.
\]
For this fixed order, replace one original item by its auxiliary variable.
At the position replaced, the two distributions face the same future
continuation values $\alg_0\ge\alg_1$ in any residual feasible state.
By \eqref{eq:excess-bellman}, the replacement can increase the expected
value by at most $\mu_i/n$. This uniform bound propagates to every earlier
position: raising each of the two future continuation values by at most
$\mu_i/n$ raises their maximum by at most that amount. Backward induction
therefore bounds the change in the value of the whole instance by
$\mu_i/n$. Replacing all marginals successively proves
\[
 \alg_\sigma\bigl((\mu_iY_{i,j_i})_i\bigr)
       \le\alg_\sigma(X)+\frac1n\sum_i\mu_i.
\]
Lemma~\ref{lem:exponential-order} applies conditional on these same indices $\{j_i\}$.
Averaging its guarantee, and using that the conditional prophet benchmarks
average to the prophet benchmark for the independent mixture distributions $\{Y_i\}$, gives
\begin{equation}\label{eq:averaged-proxy-bound}
 \opt\bigl((\mu_iY_i)_i\bigr)
 \le2\left(\bE_\sigma[\alg_\sigma(X)]+\frac1n\sum_i\mu_i\right)
 \le4\bE_\sigma[\alg_\sigma(X)].
\end{equation}
The last inequality is one of the bounds in \Cref{eq:mean-cost}, valid for every fixed order.

\emph{Unconditional prophet comparison.}
Increasing an item's excess at every threshold can only
increase the prophet benchmark. Indeed, fix all other values and write
$z_0$ for the best value without the item and $z_1$ for the best value of
other items when the item is required. Downward closure gives $z_0\ge z_1$.
For a feasible item with independent reward $W$, its conditional benchmark
is $z_0+\excess_W(z_0-z_1)$.
The claimed monotonicity follows by replacing the value distributions one at a time.
Apply it to \eqref{eq:mixture-domination}, scaled back to original units by $\mu_i$:
\begin{equation}\label{eq:shifted-prophet-domination}
 \opt\bigl((\mu_i(U_i-k)_+)_i\bigr)
    \le6k\,\opt\bigl((\mu_iY_i)_i\bigr).
\end{equation}
Independence is preserved after averaging over the sampled indices, since
every item has its own independent index and conditional reward draw.

Finally Lemma~\ref{lem:normalization-cost} bounds the cap and the removed
part below $k$ by $(2+k)\alg_\sigma(X)$ in every order. Average that bound
and combine \eqref{eq:averaged-proxy-bound} and
\eqref{eq:shifted-prophet-domination} to obtain
\[
 \opt(X)
 \le(2+k)\bE_\sigma[\alg_\sigma(X)]
       +6k\,\opt\bigl((\mu_iY_i)_i\bigr)
 \le(25k+2)\bE_\sigma[\alg_\sigma(X)].
\]
At least one of the finitely many resulting permutations has online value
at least its average. Fixing that permutation proves the last
assertion.
\end{proof}

\subsection{Construction of Auxiliary Distributions}

Now we present the actual construction of the auxiliary activated exponential variables, and show that they satisfy the desired domination conditions.

\paragraph{One finite grid for each normalized item.}
Suppress the item index and consider $0\le U\le2n$ with $\bE[U]\le1$.
Denote its ``regularized'' excess by
\[
 g(v)=\excess_U(v)+\frac1n.
\]
This function is nonincreasing, equals $1/n$ for $v\ge2n$, and satisfies
$1/n\le g(v)\le1+1/n\le2$. For any $x\le v$, therefore,
$1\le g(x)/g(v)\le n+1$.
We choose grid points $x_j=k^j$ for $0 \le j < k$, such that the $k$ intervals $[x_j,x_{j+1})$ that cover $[1,2n)$.
For notational simplicity, we also let the grid extend beyond $2n$, i.e., $x_j = k^j$ for any positive $j$.
The actual construction only involves up to $k$ intervals.

The distribution $Y_{i, j}$ ``responsible for'' interval $j$ will be based on a decreasing exponential excess curve starting at height $g(x_j)$.
Recall that we need $Y_{i, j}$ to be dominated by $U_i$, which means a sharp subsequent decline of $g$ may force the excess curve of $Y_{i, j}$ to decay quickly.
This is potentially problematic, because we also want the mixture $Y_i$ to dominate $U_i$ after shifting and scaling.
We compensate by assigning more sampling weight to that interval.
The next lemma gives the construction and both required comparisons.

\begin{lemma}[Discrete auxiliary laws]\label{lem:grid-comparison}
For every $0\le U\le2n$ with $\bE[U]\le1$, there are $k$ activated
exponential variables $Y_j$, with positive scales $s_j$, and weights $w_j>0$
such that their mixture $Y$, sampled with probabilities
$w_j/W$ where $W=\sum_jw_j$, satisfies, for every $v\ge0$,
\[
 \excess_{Y_j}(v)\le\excess_U(v)+\frac1n\quad\text{for each }j,
 \qquad
 \excess_{(U-k)_+}(v)\le\excess_{6kY}(v).
\]
All distributions $Y_j$ and weights $w_j$ depend only on the distribution of $U$ and on $n$.
\end{lemma}
\begin{proof}
Use the regularized excess $g$ and the grid $x_0, \dots, x_k$ as defined above. Write $g_j=g(x_j)$ for each interval $j$. For $0\le j < k$, set
\begin{equation}\label{eq:grid-rates-weights}
 \lambda_j=\sup_{z\ge x_j}\frac{\log(g_j/g(z))}{z},
 \qquad F_j(v)=g_j e^{-\lambda_jv},
 \qquad w_j=\left(\frac{g_j}{g_{j+2}}\right)^{1/k}.
\end{equation}
$F_j$ is an auxiliary function that simplifies presentation.
The rates $\lambda_j$ are finite because $z\ge x_j\ge1$ and $g_j/g(z)\le n+1$.
If $g_j>1/n$, evaluation at $z=2n$ also shows $\lambda_j>0$.
If $g_j=1/n$, then $U\le x_j$ almost surely; set $Y_j=0$ and $s_j=1$.
In all other cases set
\begin{equation}\label{eq:grid-activation}
 s_j=1/\lambda_j,\qquad p_j=\lambda_jg_j e^{-\lambda_j},
\end{equation}
and let $Y_j$ be an exponential of mean $s_j$ activated with probability
$p_j$. We sample the interval with probability $w_j/W$.

\emph{Each conditional distribution $Y_j$ is dominated.}
For $v\le x_j$, monotonicity gives $F_j(v)\le g_j\le g(v)$.
For $v\ge x_j$, the definition of $\lambda_j$ gives
$\lambda_jv\ge\log(g_j/g(v))$, hence $F_j(v)\le g(v)$ again.
It remains to ensure that this curve is an actual activated-exponential
excess function. Matching $F_j(v)$ without a shift would require activation
probability $\lambda_jg_j$, which need not be at most one. Instead,
\eqref{eq:grid-activation} gives
\[
 p_j\le2\lambda_j e^{-\lambda_j}\le2/e<1,
 \qquad \excess_{Y_j}(v)=F_j(v+1)\le F_j(v)\le g(v).
\]
Here $z e^{-z}\le1/e$ for $z\ge0$.
Thus the one-unit shift changes the initial height just enough to guarantee a valid probability, without changing the decay rate.

\emph{One interval pays for a target threshold.}
Fix $t\in[1,2n)$, and let $j$ satisfy $x_j\le t<x_{j+1}$.
We claim
\begin{equation}\label{eq:grid-single-interval}
 w_j F_j(t/k)\ge\tfrac12 g(t).
\end{equation}
For each $z\ge x_j$ contributing to the supremum in
\eqref{eq:grid-rates-weights}, consider the corresponding decay factor
\[
 \left(\frac{g_j}{g(z)}\right)^{t/(kz)}.
\]
There are three ranges. If $x_j\le z\le t$, then $t/(kz)\le1$ and
$g(z)\ge g(t)$, so the factor is at most $g_j/g(t)$. If
$t\le z\le x_{j+2}$, the exponent is at most $1/k$, and the factor is
at most $(g_j/g_{j+2})^{1/k}=w_j$. Finally, if $z\ge x_{j+2}$,
then $z\ge kt$ and the exponent is at most $1/k^2$. Even the full possible
decline of $g$, $n + 1$, costs at most
\[
 (n+1)^{1/k^2}\le k^{1/k}\le2.
\]
The first inequality uses $n+1\le k^k$, and the last follows from
$k\le2^k$ for integers $k\ge2$.
Since $w_j\ge1$ and $g_j/g(t)\ge1$, every factor in all three ranges is
at most $2w_jg_j/g(t)$. Taking the supremum gives
$e^{\lambda_jt/k}\le2w_jg_j/g(t)$, which is exactly
\eqref{eq:grid-single-interval}.

\emph{The weights have total at most $3k$.}
Multiplying the even-indexed weights telescopes:
\[
 w_0w_2\cdots w_{2a}
 =\left(\frac{g_0}{g_2}\frac{g_2}{g_4}\cdots
             \frac{g_{2a}}{g_{2a+2}}\right)^{1/k}
 \le(n+1)^{1/k}\le k.
\]
Here, $a=\lfloor(k-1)/2\rfloor$.
The same holds for the odd-indexed weights, and for an empty product the
bound is immediate. Since every $w_j\ge1$, expanding the product within
one parity gives
\[
 \prod_jw_j=\prod_j(1+(w_j-1))
      \ge1+\sum_j(w_j-1).
\]
Each parity therefore adds at most $k-1$ above weight one per interval.
There are at most $k$ intervals, so $W\le k+2(k-1)\le3k$.

\emph{The scaled mixture $6k Y$ covers the shifted marginal.}
For any $v\ge0$, put $t=v+k$. If $\excess_U(t)=0$, the desired inequality
is immediate. Otherwise $k\le t<2n$, so $t$ lies in a retained interval
$j$ with $g_j>1/n$. Its associated distribution $Y_j$ is nonzero, and
\[
 \excess_{Y_j}(v/k)=F_j(v/k+1)=F_j(t/k).
\]
Keep this one interval in the mixture and use
\eqref{eq:grid-single-interval} and $W\le3k$:
\[
 \excess_Y(v/k)
 \ge\frac{w_j}{W}F_j(t/k)
 \ge\frac{g(t)}{2W}
 \ge\frac{\excess_U(t)}{6k}.
\]
Because excess is nonincreasing in its threshold,
\[
 \excess_{6kY}(v)
   =6k\excess_Y\bigl(v/(6k)\bigr)
   \ge6k\excess_Y(v/k)
   \ge\excess_U(v+k)
   =\excess_{(U-k)_+}(v).
\]
This proves the second comparison and completes the lemma.
\end{proof}

\paragraph{Remark.}
The same parameter $k$ pays for two quantities: the grid has at most $k$
intervals, and each alternating product of compensating weights is at most
$k$. A target threshold $t \in [x_j, x_{j + 1})$ in the regularized excess function $g$ needs only its own interval's (i.e., $Y_j$'s) contribution. The
construction does not require an approximation by one exponential
simultaneously at all target thresholds.

Now we are ready to prove the main theorem of the section.

\begin{proof}[Proof of Theorem~\ref{thm:free-upper}]
Apply Lemma~\ref{lem:grid-comparison} independently to every normalized
capped variable $U_i$, using its weights to define $\Daux_i$. The resulting
distributions satisfy \eqref{eq:conditional-domination} and
\eqref{eq:mixture-domination}. Lemma~\ref{lem:excess-reduction} then gives
$\opt(X)\le(25k+2)\bE_\sigma[\alg_\sigma(X)]$ and hence the stated guarantee
for some fixed order.
For $k>2$, minimality gives $(k-1)\log(k-1)<\log(2n+2)\le k\log k$; these inequalities imply $k=\Theta(\log n/\log\log n)$ as $n$ grows. This proves the asserted
asymptotic ratio.
\end{proof}

\section*{AI Disclosure}
\addcontentsline{toc}{section}{AI Disclosure}

The author interacted with GPT-5.6 and GPT-6 extensively and adaptively in the process of obtaining the results.
All proofs were digested, verified, simplified, and reinterpreted by the author.
In particular, the layered self-encoding interpretation, the earliest divergence formulation in the lower bound for the standard prophet problem, and the harvesting game in the lower bound for the prophet secretary problem are human.
The author also persistently instructed GPT-6 to find a discrete construction of the upper bound for the free-order prophet problem, which is (in the author's opinion) significantly cleaner than the continuous sampling argument initially found.
The author used GPT-6 to assist in writing, mostly in the main technical sections.
All sections were verified and polished by the author, who takes full responsibility of all contents.

\addcontentsline{toc}{section}{References}
\bibliographystyle{plainnat}
\bibliography{ref}

\end{document}